\documentclass{agujournal2019}
\usepackage{url} 
\usepackage{soul}
\usepackage[utf8x]{inputenc}
\usepackage[T1]{fontenc}
\usepackage{amsmath}
\journalname{JGR: Planets}

\begin{document}

%
%


\title{Thermal structure and aerosols in Mars' atmosphere from TIRVIM/ACS onboard the ExoMars Trace Gas Orbiter :  validation of the retrieval algorithm}

%
%




\authors{S. Guerlet\affil{1,3}\thanks{4 Place Jussieu, Paris, France}, N. Ignatiev\affil{2}, F. Forget\affil{1}, T. Fouchet\affil{3},  P. Vlasov\affil{2}, G. Bergeron\affil{1}, R. M. B. Young\affil{1,4}, E. Millour\affil{1}, S. Fan\affil{1}, H. Tran\affil{1}, A. Shakun\affil{2}, A. Grigoriev\affil{2}, A. Trokhimovskiy\affil{2}, F. Montmessin\affil{5} \ and O. Korablev\affil{2}}


\affiliation{1}{Laboratoire de Météorologie Dynamique (LMD/IPSL), Sorbonne Université, ENS, PSL Research University, Ecole Polytechnique, Institut Polytechnique de Paris, CNRS, Paris, France}
\affiliation{2}{Space Research Institute (IKI), Moscow, Russia}
\affiliation{3}{LESIA, Observatoire de Paris, Université PSL, CNRS, Sorbonne Université, Université de Paris, 5 place Jules Janssen, 92195 Meudon, France}
\affiliation{4}{Department of Physics \& National Space Science and Technology Center, United Arab Emirates University, Al Ain, UAE}
\affiliation{5}{LATMOS/IPSL, Guyancourt, France}





\correspondingauthor{Sandrine Guerlet}{sandrine.guerlet@lmd.jussieu.fr}




\begin{keypoints}
\item We exploit TIRVIM spectra to determine Martian atmospheric, surface temperature, as well as integrated opacity of dust and water ice clouds.
\item Different sources of biases are investigated with the help of simulated observations at different local times, latitudes and seasons.
\item Atmospheric temperatures retrieved from TIRVIM in April-May 2018 are in excellent agreement with co-located MCS observations.
\end{keypoints}

%
%


\begin{abstract}

The Atmospheric Chemistry Suite (ACS) onboard the ExoMars Trace Gas Orbiter (TGO) monitors the Martian atmosphere through different spectral intervals in the infrared light. We present a retrieval algorithm tailored to the analysis of spectra acquired in nadir geometry by TIRVIM, the thermal infrared channel of ACS. Our algorithm simultaneously retrieves vertical profile of atmospheric temperature up to 50 km, surface temperature, and integrated optical depth of dust and water ice clouds. The specificity of the TIRVIM dataset lies in its capacity to resolve the diurnal cycle over a 54 sol period. However, it is uncertain to what extent can the desired atmospheric quantities  be accurately estimated at different times of day. Here we first present an Observing System Simulation Experiment (OSSE). We produce synthetic observations at various latitudes, seasons and local times and run our retrieval algorithm on these synthetic data, to evaluate its robustness. Different sources of biases are documented, in particular regarding aerosol retrievals. Atmospheric temperature retrievals are found robust even when dust and/or water ice cloud opacities are not well estimated in our OSSE. We then apply our algorithm to TIRVIM observations in April-May, 2018 and perform a cross-validation of retrieved atmospheric temperature and dust integrated opacity by comparisons with thousands of co-located Mars Climate Sounder (MCS) retrievals. Most differences between TIRVIM and MCS atmospheric temperatures can be attributed to differences in vertical sensitivity. Daytime dust opacities agree well with each other, while biases are found in nighttime dust opacity retrieved from TIRVIM at this season.

\end{abstract}

%
%

\section*{Plain Language Summary}
The Martian surface and atmosphere undergo strong variations in temperature and amount of aerosols (dust or water ice cloud particles). Our knowledge on their variations at diurnal scale is however limited, due to lack of appropriate observations. 
We present a method to analyze thermal emission spectra of Mars’ surface and atmosphere recorded by TIRVIM, a spectrometer onboard the ExoMars Trace Gas Orbiter.
We have developed a program to derive surface and atmospheric temperatures from these spectra, along with an estimation of the amount of aerosols. The specificity of the TIRVIM dataset is its capacity to resolve the diurnal cycle over a 54 sol period.  However,  atmospheric quantities cannot be accurately estimated at all times of day. One of the goals of our paper is to assess the robustness of our algorithm with the help of simulated observations. The retrieval of aerosol opacity is assessed to be challenging at some times of day, but atmospheric temperature is well determined. We have then applied our algorithm to tens of thousands of TIRVIM observations obtained in April-May 2018 and showed that our derived atmospheric temperatures compare very well with independent measurements obtained from the Mars Climate Sounder, reinforcing our confidence in our method.


\section{Introduction \label{sec:intro}}

The ESA-Roscosmos ExoMars Trace Gas Orbiter (TGO) was launched on March 14th, 2016 and was successfully inserted into Mars' orbit in November, 2016. After several orbit manoeuvres including a 12-month aerobraking phase it reached its final, 400-km altitude and near-circular orbit on April 13th, 2018.
Onboard TGO, the Atmospheric Chemistry Suite (ACS) comprises three spectrometers, each tailored for specific scientific goals \cite{Korablev2018}. 
We focus here on the thermal infrared spectrometer, named TIRVIM (Thermal InfraRed channel in honor of professor Vassilii Ivanovich Moroz), which was operational from April, 2018 until December, 2019, hence almost a martian year (between mid- martian years 34 and 35). 
Through nadir-viewing soundings, its main goal is to monitor the atmospheric temperature, surface temperature and integrated aerosol content -- dust and water ice clouds -- at a great variety of local times. 
Indeed, TGO's orbit is designed in such a way that nadir observations sample the full diurnal cycle in 54 sols, at all latitudes between 74°N and 74°S. Hence, it is able to capture both seasonal and diurnal variations of these climatological variables and complements  other still-operating thermal infrared sounders, such as the Mars Climate Sounder (MCS) onboard NASA's Mars Reconnaissance Orbiter, the Planetary Fourier Spectrometer (PFS) onboard ESA's Mars Express and the Thermal Emission Imaging System (THEMIS) onboard Mars Odyssey.

MCS is a radiometer operating in limb-viewing geometry, allowing the retrieval of vertical profiles of atmospheric temperature from 5 to 80~km as well as dust and water ice vertical profiles, with a vertical resolution of typically 5~km \cite{Kleinbohl2009}. Being on a Sun-synchronous polar orbit, MCS mostly acquires data at local times 3:00 and 15:00 ($\pm$ 1.5 hours when cross-track observations are performed).
PFS is a thermal infrared spectrometer operating in nadir-viewing geometry providing temperature profiles in the range 5--50~km with a vertical resolution of 10~km \cite{Grassi2005} along with surface temperature and the integrated content of dust and water ice clouds. The same quantities can be retrieved from TIRVIM observations. The Mars Express orbiter is such that PFS observations sample various local times but with a longer revisit time compared to TIRVIM (150 sols for PFS versus 54 sols for TIRVIM), and a sparser spatial coverage due to the elliptical orbit of Mars Express.
Finally, the THEMIS instrument comprises several cameras that image Mars in the visible and thermal infrared. It is sensitive to atmospheric temperature in a broad altitude range centered on 50~Pa, and allows for the retrieval of integrated dust and water ice opacity \cite{Smith2009}. It is on a Sun-synchronous orbit, but the local time coverage has varied over the past 20 years; it sampled local times 7AM--7PM during Martian Year 34.
Another important instrument to add to this list was the Thermal Emission Spectrometer (TES) mounted on the Mars Global Surveyor that operated between 1997 and 2006. It was a spectrometer rather similar to PFS with a coarser spectral resolution of either 5 or 10~cm$^{-1}$, sensitive to the temperature in the range 5--35~km, dust and water ice cloud opacities in nadir mode \cite{Conrath2000, Smith2000}. It performed measurements at local times near 2~AM and 2~PM and also operated systematically in limb geometry, allowing the retrieval of atmospheric temperature up to 65 km.

Deriving those atmospheric quantities is of high interest to study the Martian climate at various spatial and temporal scales, from diurnal variations to inter-annual variations. 
These measurements have helped to broadly characterize  the vertical and meridional structure of atmospheric temperature, dust loading and water ice clouds
\cite<e.g.,>[]{Smith2001,McCleese2010, Giuranna2021},
which in turn bring insights onto Mars atmospheric dynamics. In particular, the TES and MCS retrieval products have been used in several data assimilation studies \cite<e.g.,>[]{Steele2014, Navarro2014, Greybush2019}.
More recently, these thermal infrared measurements have brought new insights on the diurnal variability of dust and water ice clouds \cite{Kleinbohl2020, Smith2019Icar, Wolkenberg2021, Giuranna2021}. 
However, developing a retrieval algorithm that performs well at all conditions (dust load, local time, etc) is a challenge. 
Particular care has to be taken regarding the reliability of aerosol retrievals from nadir sounders when the contrast between the surface temperature and atmospheric layer where the aerosols lie is low, as  already raised by several previous studies  \cite{Pankine2013, Smith2019Icar,SmithDust2019, Wolkenberg2021}.

The objective of this paper is to document a retrieval algorithm developed at Laboratoire de M\'et\'eorologie Dynamique (LMD) for the analysis of TIRVIM observations, to discuss some challenges identified and validate our retrievals against independent observations from MCS. 
We first detail characteristics of TIRVIM observations in section~\ref{sec:obs}.
We describe our algorithm in section~\ref{sec:algo} and apply it to synthetic measurements generated for a great variety of scenes (latitudes, local time, surface temperature, aerosol content, topography) in section~\ref{sec:synthe}. Challenges arise due to the degeneracy in the inverse problem, especially regarding  the combined retrieval of surface temperature, dust and water ice optical depth. 
This algorithm is then applied to the first 45 sols of TIRVIM observations from mid-March to end of April 2018, as described in section~\ref{sec:results}. 
The retrieved temperature profiles are compared to co-located MCS observations near 3~AM and 3~PM to perform a cross-validation of our retrievals and evaluate potential biases.
A first assessment of the quality of dust retrievals is also included.
Detailed discussion regarding the diurnal temperature variations derived from TIRVIM observations is deferred to another paper. We conclude on the performance of TIRVIM and on our algorithm in section~\ref{sec:discuss}.


\section{ACS/TIRVIM nadir measurements \label{sec:obs}}
\subsection{Instrument characteristics}

TIRVIM is a double-pendulum Fourier-transform spectrometer sensitive in the spectral range 1.7 to 17 $\mu$m (600 -- 5200~cm$^{-1}$). 
It operates routinely in nadir-viewing geometry and can also operate in solar occultation mode; the former type is the focus of this paper. In nadir geometry, spectra are only exploitable between 620 and 1300~cm$^{-1}$, as thermal emission from the surface and atmosphere quickly drops beyond 1000~cm$^{-1}$.
In this geometry, the apodized spectral resolution is 1.2~cm$^{-1}$ and the integration time for a single interferogram is 0.4 seconds.
The individual projected field of view on Mars from TGO's circular orbit for a single observation is 14~km cross-track $\times$~16~km along-track, accounting for the small (4~km/s) smearing due to the spacecraft's motion during acquisition.
Further details on TIRVIM technical characteristics can be found in ~\citeA{Korablev2018}.

\subsection{Calibration and instrumental issues} 

Generation of calibrated spectra from raw interferograms is done at the Space Research Institute (IKI) in Moscow, Russia. Absolute radiometric calibration was facilitated by routine periodical measurements of the internal calibration black body and of the cold space. TIRVIM turned out to be an IR Fourier transform spectrometer with a source-dependent phase function. Radiometric calibration of such an instrument was considered by \citeA{Revercomb1988}.

\begin{figure}
  \centering
  \includegraphics[width=0.75\linewidth]{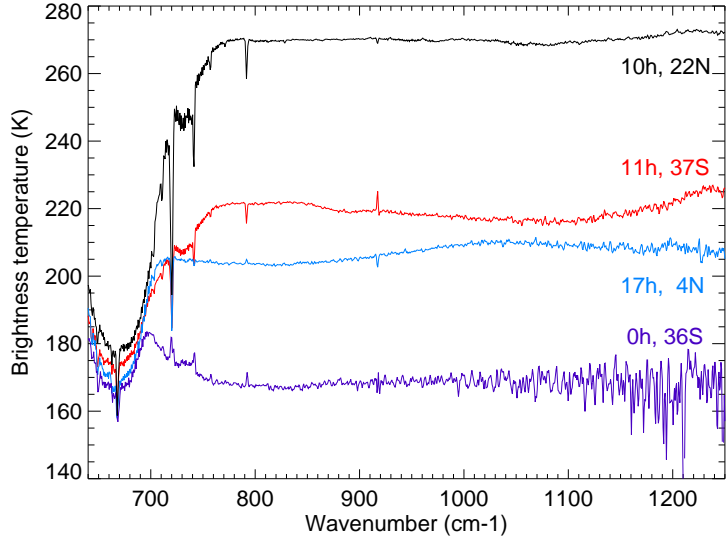}
  \caption{Examples of TIRVIM spectra acquired in March 2018 at four latitudes and local times, with different surface temperatures. They were acquired in the onboard averaging mode, where 8 interferograms were averaged by the instrument software onboard TIRVIM. Absorption by dust centered at 1090~cm$^{-1}$ is particularly visible in the spectrum at 37°S, 11h (in red), while absorption by water ice clouds  centered at 820~cm$^{-1}$ is visible in the spectrum acquired at 4°N, 17h (in blue). An electric spike, which is an artifact, can be seen near 920~cm$^{-1}$.}
  \label{fig:ex_spectra}
\end{figure}

Ahead of the beginning of TIRVIM science operations, the chosen strategy to relay the data was to perform an on-board averaging of 8 consecutive interferograms in order to increase the signal-to-noise ratio (SNR). This was justified by the fact that the individual interferograms are recorded at very close locations on Mars, hence surface and atmospheric variables should not vary much amongst 8 sequential acquisitions. 
This on-board averaging mode was used for the first five months of the mission (except for the very first six days, where individual interferograms were relayed).
However, the onboard interferogram averaging appeared to be sometimes incorrect. Interferograms must be aligned in the path difference space before averaging. The onboard TIRVIM software aligned interferograms before averaging according to their maxima. Since the source was changing, and the instrument had a source dependent phase, maxima of the interferograms did not always correspond to the same path difference. In such cases, the average interferogram and therefore calibrated spectral radiance was incorrect. 
Those cases were flagged as  poor-quality data and were excluded from our analysis. They could represent up to 20\% of all measurements. Fortunately, the other $\sim$80\% are of good quality and are exploited in this paper. The corresponding field of view of the averaged interferograms is approximately 25$\times$105~km. 
To limit data loss, this mode was later abandoned in favor of recording and relaying each individual interferograms. The drawback was the reduced signal-to-noise ratio (down to 3 times) and a higher data rate from the spacecraft.
Among the other known issues in the calibrated spectra is an electrical spike near 920~cm$^{-1}$ (that is excluded from our analysis).  

Examples of TIRVIM spectra acquired in March 2018 at different local times and latitudes are shown in units of brightness temperature in Figure~\ref{fig:ex_spectra}. 
Apart from the near blackbody surface emission, these spectra are dominated by the atmospheric CO$_2$ absorption band centered at 667~cm$^{-1}$ (sometimes visible in emission, when the atmosphere is warmer than the surface), a broad water ice feature centered at 820~cm$^{-1}$ and a broad dust feature centered at 1090~cm$^{-1}$.
We emphasize here that the left wing of the broad CO$_2$ absorption band is almost not captured by TIRVIM (unlike PFS). Apart from a few particular cases (mentioned later), this does not hamper the atmospheric temperature retrievals.

\begin{figure}
  \centering
  \includegraphics[width=0.75\linewidth]{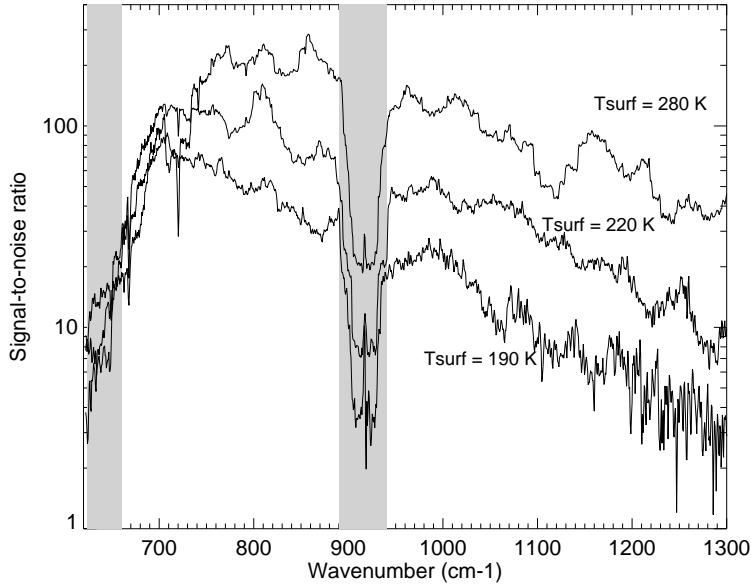}
  \caption{Signal-to-noise ratio for three individual spectra (no onboard averaging) acquired in October 2018, for which the retrieved surface temperature was 190K, 220K and 280K, as labeled. Note that the SNR is similar for the three spectra in the range 660--700~cm$^{-1}$, as in this spectral range, the outgoing thermal radiation depends on the atmospheric temperature and not on surface temperature. The two grey shaded areas have large noise levels (low SNR) while not being essential to our analysis and are thus excluded. One corresponds to the left edge of the spectrum (620--660~cm$^{-1}$), the other one to an electric spike centered at 920~cm$^{-1}$.}
  \label{fig:snr}
\end{figure}

The Noise Equivalent Radiance (NER) of a calibrated spectrum obtained without interferogram averaging is of the order of 0.4$\times 10^{-7}$~W~cm$^{-2}$sr$^{-1}$/cm$^{-1}$ for most of the spectral range, and increases near the edge of our spectral domain: it reaches 1.2$\times 10^{-7}$~W~cm$^{-2}$sr$^{-1}$/cm$^{-1}$ at 660~cm$^{-1}$ and  0.8$\times 10^{-7}$~W~cm$^{-2}$sr$^{-1}$/cm$^{-1}$ at 1300~cm$^{-1}$. These values are comparable to the NER of PFS spectra at 600~cm$^{-1}$ but are up to 10 times lower than PFS NER at 1300~cm$^{-1}$  \cite{Giuranna2005}.
The resulting signal-to-noise ratio strongly depends on the Martian surface (or atmospheric) temperature and wavenumber. Figure~\ref{fig:snr} illustrates these SNR variations: for a warm surface (eg. 280K) the SNR is in the range 30--200, depending on wavenumber, while for a colder surface (eg. 190K), the SNR is typically ten times lower.
An exception is for wavenumbers 650--700~cm$^{-1}$, where surface emission does not contribute to the measured thermal emission from space and for which the SNR depends on atmospheric temperature, which varies less dramatically than the surface temperature.
In this wavenumber range, the SNR is of the order of 30--100.
If we consider spectra resulting from the onboard averaging of 8 interferograms, the aforementioned SNR values are multiplied by $\sim$3.

In this paper, we will include analysis of spectra acquired in both modes, predominantly the averaging mode.
We will see that good performances are achieved with this mode.
The calibration version we use is referred to as version 4, where  orbit-average black body and space interferograms are used for calibration.

\subsection{Spatiotemporal coverage of TIRVIM nadir measurements}

\begin{figure}
  \centering
  \includegraphics[width=0.95\linewidth]{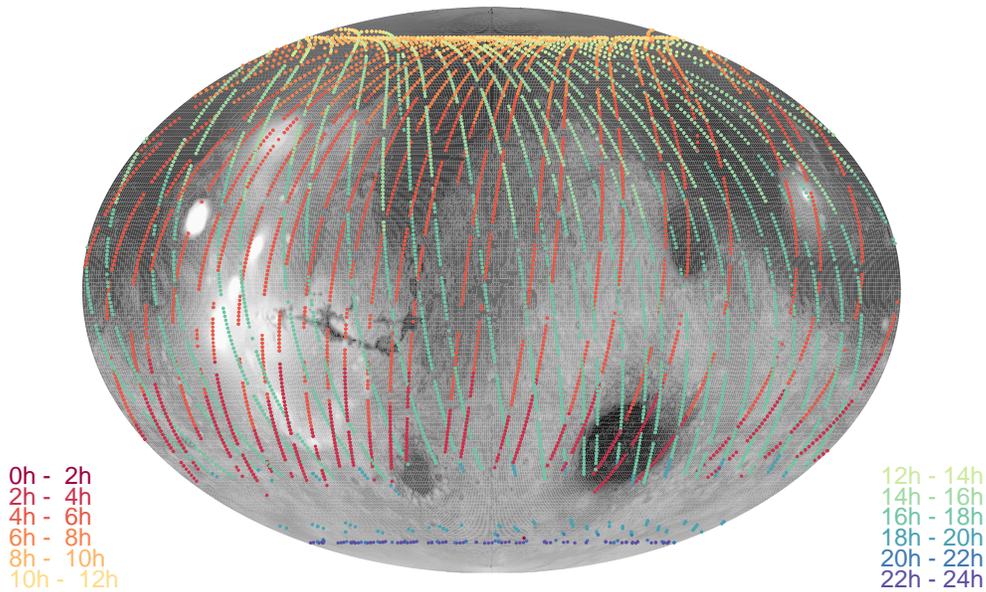}
  \caption{Coverage of TIRVIM nadir observations acquired on the 26th, 27th and 28th of March 2018, showing here only data that passed a set of initial quality filters. Different colors refer to the local time, as labeled. TGO's orbit is such that the local time of TIRVIM observations shifts  by $\sim$13~minutes earlier every sol. }
  \label{fig:coverage_mola}
\end{figure}

The ExoMars Trace Gas Orbiter is set on a near-circular orbit at 400-km altitude with an inclination of 74° (implying that nadir observations cannot be made at latitudes poleward of 74°). TGO executes 12 orbits per (Earth) day, hence sampling 24 different longitudes per day at a given latitude. 
An example of the coverage obtained after three days (36 orbits of TGO) is shown in Figure~\ref{fig:coverage_mola}. At low latitudes, on this short time period, mostly two local times are sampled. 
However, TGO is not in a Sun-synchronous configuration: rather, the local time coverage slightly drifts earlier each day in such a way that after 54 sols (corresponding to 25°--35° of Ls, depending on season), the diurnal cycle has been evenly sampled in local times over the whole planet -- providing that TIRVIM is operating continuously.  
The revisit time for a given (latitude, longitude, local time) targeted point on Mars is actually 108 sols ; however, if we consider an area of $\sim$5°$\times$5°, a 54-sol observation period provides coverage at all local times and is relevant for studying diurnal variations (with the caveat that $\sim$30° of solar longitude has passed).

A Stirling cryo-cooler ensured the stability of the detector operating temperature at 65--75K. Several overheating events occurred in the first months of the mission, which required the cooler to be switched off for several weeks, which meant the absence of meaningful data.
TIRVIM started its routine nadir observations on March 13, 2018 but encountered an issue with the stability controller of the double pendulum movement on April 28, 2018, which caused dramatic loss of the data quality. It resumed its quality observations on May 26, 2018 until an overheating event occurred on July 15, 2018. 
After that, overheating events became more frequent. To overcome these issues, from September, 2018 onwards, an automatic switch off of the cryo-cooler was programmed if its temperature exceeded 14°C.
Another limitation stemmed from the lifetime of the Stirling cryo-cooler, which was estimated to last 10 000 hours (it actually operated 8,000~hours before failing in December, 2019). If TIRVIM had been switched  ON all the time, then the cooler would have stopped functioning after 10 months. In order to mitigate this effect, a duty cycle of $\sim$50\% was undertaken in October, 2018 at which time TIRVIM acquired data roughly 2 days out of 7, except for one 9-day long acquisition every month. This observing strategy largely prevented overheating events from occurring, and allowed TIRVIM to acquire data over almost a full martian year.
In this paper, we will report in section~\ref{sec:results} on the analysis of the first 45 days of TIRVIM data during March-April, 2018. 


\section{Retrieval algorithm \label{sec:algo}}

TIRVIM spectra contain information on the surface temperature, on the vertical profile of atmospheric  temperature  between a few kilometers above the surface and $\sim$50~km (or 2--3~Pa) and on the column integrated opacity of dust and water ice clouds.
We aim at retrieving simultaneously these quantities by exploiting TIRVIM spectra between 660 and 1300~cm$^{-1}$.
Our algorithm comprises a forward radiative transfer model used to generate synthetic observations, coupled with a constrained linear inverse model, described below. 

\subsection{Radiative transfer model}

Our forward radiative transfer model computes the spectral radiance $I_\nu$ of the outgoing thermal emission of Mars' surface and atmosphere, neglecting scattering contributions, in the plane-parallel approximation :

\begin{equation}
    I_\nu (\tau=0, \mu)= \epsilon_\nu B_\nu (T_\mathrm{surf}) e^{-\tau_\mathrm{total}/ \mu} + \frac{1}{\mu} \int_0^{\tau_\mathrm{total}} B_\nu(T(\tau')) e^{-\tau' / \mu} d\tau'
    \label{eq:rad}
\end{equation}

\noindent with $\mu=1/\mathrm{cos}(\theta)$ the airmass at an emission angle $\theta$ ; $\epsilon_\nu$ the surface emissivity; $B_\nu$ the Planck function, $\tau'$ the partial integrated optical depth from the top of the atmosphere to a given pressure level, $T_\mathrm{surf}$ the surface temperature and $T$ the atmospheric temperature at a given pressure level. 
Our model atmosphere is discretized into 45 vertical sigma-levels, with the first level just above the surface. Hence, rather than using a fixed pressure grid, our pressure grid is tuned to local surface pressure to adapt to the topography.
Radiances from TIRVIM and those calculated from equation~\ref{eq:rad} are converted in brightness temperatures  $T_{B,\nu}$. Unless stated otherwise, in the following, we work in brightness temperature units and not radiances.

\begin{figure}
  \centering
  \includegraphics[width=0.75\linewidth]{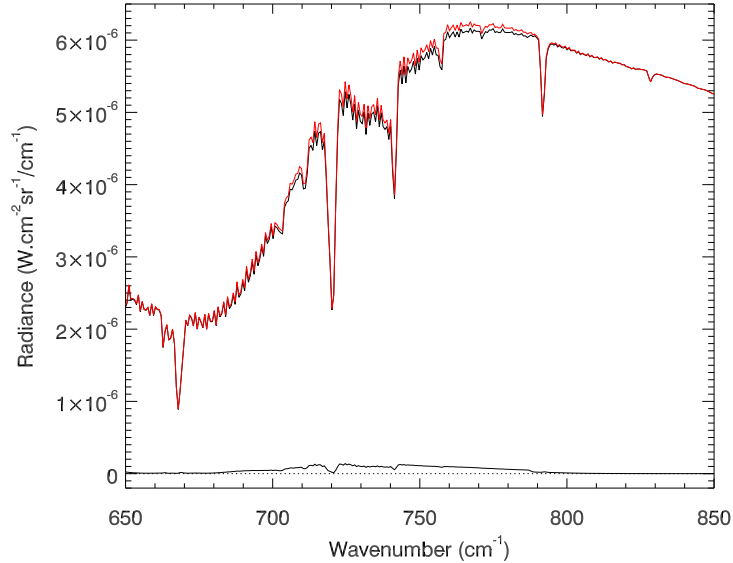}
  \caption{Synthetic TIRVIM spectra focusing here on the CO$_2$ band, assuming (in black) a Lorentz line profile or (in red) a sub-Lorentzian line profile to empirically take into account the effects of collisional line mixing and the finite duration of
collisions, based on experimental work by \citeA{Perrin1989}. The difference between the two spectra is plotted at the bottom.}
  \label{fig:co2_sublo}
\end{figure}

We use a line-by-line approach, where spectra are first computed at a high spectral resolution of 0.01~cm$^{-1}$, and are then convolved at the resolution of the instrument. TIRVIM spectra are apodized with a Hamming function and we use the appropriate instrument function for this convolution. Note that the sampling is 0.645~cm$^{-1}$ and the corresponding spectral resolution (full width at half maximum) is of 1.2~cm$^{-1}$.
We take into account opacity from atmospheric CO$_2$ gas, dust and water ice clouds.
Line-by-line, CO$_2$ absorption coefficients are tabulated offline for a set of 45 reference pressures equally spaced in natural logarithm space (from 1260~Pa to 5$\times$10$^{-3}$~Pa) and 12 temperatures. These sets of 12 reference temperature values themselves depend on pressure and are sampled to encompass climatological conditions over a full martian year, including global dust storm conditions with warmer temperatures. For instance, near the surface (p$>$ 300 Pa), reference temperature values range from 150K to 260K, every 10K while at 10 Pa, these levels comprise ten levels sampled from 100 to 190K (every 10K) plus two extra levels at 210K and 230K.
These computations use the HITRAN 2016 spectroscopic database \cite{HITRAN2016}. We tested another set of CO$_2$ coefficients generated from the GEISA 2015 linelist \cite{GEISA2015} and found that it yielded similar spectra, as far as the CO$_2$ 15~$\mu$m band is concerned. 
The temperature dependence of the line width due to pressure broadening is not taken from the HITRAN database (as it corresponds to air broadening) but instead is adapted for a CO$_2$ atmosphere. It is computed for each transition, as a function of the rotational quantum number $J$, based on values tabulated by \citeA{lamouroux2012}.
Furthermore, pressure broadening is not represented by a standard Lorentz function. Rather, we adopt an
asymmetric sub-Lorentzian profile, which empirically takes into account
the effects of collisional line mixing and the finite duration of
collisions. 
These line shapes are represented by a Lorentz function multiplied by a factor $\chi$ that depends on the distance from the line center and was derived from experimental work relevant for the 4.3-$\mu$m CO$_2$ band \cite{Perrin1989}. We assume that these $\chi$ factors can also be used for the 15-$\mu$m CO$_2$ band. 
The final line shape is obtained by a convolution of the sub-Lorentzian profile with a Gaussian profile (that represents Doppler broadening of the lines).
An example of synthetic TIRVIM spectra is shown in Figure~\ref{fig:co2_sublo}, which compares spectra computed with a Lorentz function or the sub-Lorentzian line profile.
In this example, the radiance is increased by 1--2~\% (hence, 1 to 2-$\sigma$ above the noise level) in the range 700--780~cm$^{-1}$ when a sub-Lorentzian line profile is adopted.
This effect is noticeable, but its influence is small regarding atmospheric temperature retrievals(typically $\sim$ 1--2 K).
 
Extinction coefficients $Q_\mathrm{ext}$ for water ice particles are computed offline using Mie theory, assuming a log-normal size distribution, an effective radius $r_\mathrm{eff}$ of 1.45~$\mu$m, a variance of 0.1 and optical constants from \citeA{Warren1984}.
Particle sizes of typically 1 to 3.5~$\mu$m are most frequently observed, as determined from CRISM observations \cite{Guzewich2019} or from MGS TES \cite{Clancy2003}. 
Assuming a single particle size for water ice clouds in our radiative transfer model should not affect our retrievals of the cloud integrated optical depth because it is sensitive to the ratio $Q_\mathrm{ext}/r_\mathrm{eff}$. Indeed, the latter quantity varies by less than 10\% for particles sizes in the range 1--3.5~$\mu$m.
Hence, the uncertainty on total cloud opacity related to a wrong assumption of particle sizes should be on the order of 10\%.

Regarding dust particles, their extinction coefficients are computed from a T-matrix code using an effective radius $r_\mathrm{eff}$ of 1.5~$\mu$m, a variance of 0.3 and optical constants derived from CRISM observations by  \citeA{Wolff2009}, which also used a T-matrix code in the analysis.
Particle effective sizes of 1--2~$\mu$m with a variance of the size distribution of 0.2--0.4 are the most frequently observed (see review by \citeA{Khare2017} and recent results from  solar occultations recorded by ACS and analysed by \citeA{Luginin2020}). Changing the dust particle radius to 1~$\mu$m or 2~$\mu$m affects the ratio $Q_\mathrm{ext}/r_\mathrm{eff}$ by less than 10\%, hence the uncertainty on total dust opacity linked to assuming a constant dust particle radius of 1.5~$\mu$m remains small as well ($<$ 10\%).

Dust is assumed to be well-mixed (ie., with a constant mass mixing ratio) in the first two scale heights above the surface, then its mixing ratio decreases linearly with height (log-pressure, to be precise).
The vertical profile of ice mixing ratio is parameterized by a Gaussian profile centred at a condensation level that can either be set arbitrarily, or is computed by combining knowledge on the temperature profile (being retrieved simultaneously) and the water vapor column, taken from the Thermal Emission Spectrometer (TES) climatology \cite{Smith2004}.
More information on the choice for \textit{a priori} parameters for dust and water ice clouds is given in section~\ref{sec:inv:aero}.
As we neglect scattering effects, our retrieved quantities are "effective" dust and water ice absorption. The error induced by neglecting scattering was previously estimated by \citeA{Smith2004}. The authors ran tests including or not scattering effects and concluded that the actual extinction optical depth was approximately 1.3 and 1.5 times as large as the “effective” absorption optical depth for dust and water ice, respectively. Their work was based on the same dust and ice features as ours, i.e. at 1075 and 825 cm$^{-1}$, and similar particle sizes. Similar factors between absorption and extinction values were found by \citeA{Wolff2003}.

Additional ancillary data are needed to compute synthetic spectra. For each TIRVIM measurement, surface pressure is extracted at the corresponding season, local time and location from the Mars Climate Database (MCD) version 5.3 \cite{Millour2018} using the surface pressure predictor described in \citeA{Forget2007}.
The latter exploit high resolution MOLA (32 pixels/degree) topography. 
The CO$_2$ mixing ratio vertical profile is also extracted from the MCD.
Spectral surface emissivity for each observation is interpolated from TES spectral emissivity map \cite{Bandfield2003}. As information  within the CO$_2$ band is missing, we adopt a simple linear interpolation of surface emissivity in this spectral range. An error in surface emissivity in this region should not impact our results, as the contribution from the surface to the outgoing thermal emission is negligible in this spectral range of strong CO$_2$ absorption.

\subsection{Retrieval algorithm}

The overall goal is to find a set of parameters (vertical profile of temperature, surface temperature, integrated opacity of aerosols) that produces a synthetic spectrum (using the aforementioned forward radiative transfer model) in close agreement with a TIRVIM spectrum, within noise levels.
As we are facing an ill-posed and underconstrained inverse problem, there exists a strong degeneracy of this set of parameters, including potentially non-physical solutions that could still match the observed spectra. 
In order to regularize the inverse problem, we choose the widely-used framework of optimal estimation retrieval, described in \citeA{Rodgers2000}.
In this framework, the cost function to be minimized includes not only the evaluation of goodness-of-fit to the data ($\chi^2$), but also an additional regularization term that should contain our best \textit{a priori} physical knowledge on the desired quantities. To be more accurate, we use the same choice of cost function as \citeA{Conrath2000} that is slightly different from that described in \citeA{Rodgers2000} (see next subsection). 
The choice of the \textit{a priori} values together with their covariance matrix is not trivial, though, requiring a certain level of trial-and-error process.

We detail below the retrieval scheme for atmospheric temperature on the one hand, surface temperature and aerosols on the other hand, and will lastly describe how they are combined with a detailed description of the different steps of our algorithm.

\subsubsection{Atmospheric temperature retrieval scheme}
Regarding the retrieval of temperature vertical profiles, we adopt the same approach as \citeA{Conrath2000} for the analysis of the TES/MGS data, also successfully applied to PFS/Mars Express data by \citeA{Fouchet2007}. 
A first guess temperature profile $T_a$ is iteratively modified following this equation:

\begin{equation}
\label{eqT1}
    \mathbf{T}_{n+1} = \mathbf{T_a} +  \mathbf{W} (\mathbf{\Delta T_b} + \mathbf{K} (\mathbf{T}_n - \mathbf{T_a}))
\end{equation}

\begin{equation}
\label{eqT2}
    \mathbf{W}= \mathbf{S_a K^T} ( \mathbf{KS_aK^T}+ \frac{1}{\alpha}\mathbf{S_e})^{-1}
\end{equation}

where $\mathbf{T}_n$ is the temperature profile at iteration $n$, $\mathbf{\Delta T_b}$ the difference between the synthetic spectrum (in brightness temperature) computed with temperature $\mathbf{T}_n$ and the TIRVIM one, $\mathbf{K}$ is the functional derivative matrix, with elements $K_\mathrm{ij}$ defined as the derivative of the brightness temperature at wavenumber $i$ over the temperature at a pressure level $j$ ($K_\mathrm{ij}=\frac{dI_i}{dT_j} $), $\mathbf{S_a}$ the covariance matrix for the \textit{a priori} temperature profile, and $\mathbf{S_e}$ the error covariance matrix, whose diagonal elements contain the squared wavenumber-dependent NER of TIRVIM spectra. Finally, $\alpha$ is a parameter used to assign more or less weight to the observations with respect to our \textit{a priori} knowledge.
Based on several tests on retrievals on both synthetic and actual TIRVIM data, we find that a value of $\alpha= \beta \times \frac{trace(\mathbf{S_e})}{trace(\mathbf{KS_aK^T})}$, with a nominal value of 3 for $\beta$, yields  satisfactory results.

\subsubsection{Information content for the temperature}

Only a portion of a TIRVIM spectrum is exploited in equations~\ref{eqT1} and~\ref{eqT2}, as information on the atmospheric temperature comes from the CO$_2$ band in the range 660--780~cm$^{-1}$ (recall that we exclude data in the range 620--660~cm$^{-1}$ due to low signal-to-noise ratio).
Actually, within this spectral range, there exists a strong redundancy in the information content, where contribution functions at different wavenumbers peak at the same pressure level.
While some level  of redundancy is desirable (because of noise in the data), all individual spectral measurements within this range are not needed to achieve a satisfactory temperature profile retrieval.
Hence, in order to reduce computation time, we select 50 out of the 177 wavenumbers in TIRVIM spectra in the range 660--780~cm$^{-1}$. This reduces the matrix size of $\mathbf{K}$, $\mathbf{S_e}$ and $\mathbf{\Delta T_b}$ in equations~\ref{eqT1} and~\ref{eqT2}.
We keep all 14 points between 665.3~cm$^{-1}$ and 673.7~cm$^{-1}$, then one every three points until 714.3~cm$^{-1}$, then one every five points up to 780~cm$^{-1}$. 
We emphasize that we have tested retrievals in different configurations (50 or 177 points in the CO$_2$ band) and confirm that retrieved temperature profiles are almost identical.

\begin{figure}
  \centering
  \includegraphics[width=0.6\linewidth]{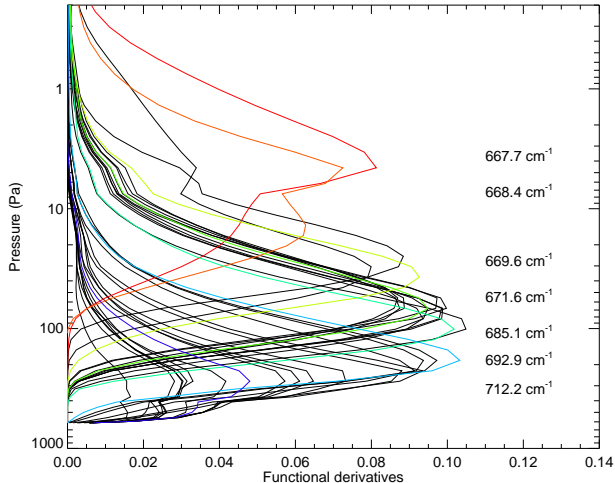}
  \caption{Functional derivatives of the temperature as a function of pressure, for fifty different wavenumbers in the range 665--780~cm$^{-1}$. Seven out of fifty wavenumbers are highlighted in color and labeled. In general, TIRVIM spectra probe from lower to higher pressure regions starting from the core of the CO$_2$ band (at 667.7~cm$^{-1}$) to its wing. This example was generated for atmospheric conditions extracted from the Mars Climate Database for Martian Year 25, Ls=180°, latitude 15°N and local time 9~AM.}
  \label{fig:ex_kk}
\end{figure}

We present an example of the information content sampled by these 50 wavenumbers, as a function of pressure, in Figure~\ref{fig:ex_kk} (examples at other latitudes and local times will be given in  section~\ref{sec:synthe}, focused on synthetic retrievals).
The exact pressure levels at which the functional derivatives peak and their relative amplitudes depend on the temperature profile itself and on the aerosol load ; however, general trends can be drawn. The radiance or brightness temperature at wavenumbers 670--780~cm$^{-1}$ is mostly sensitive to the atmospheric temperature at altitudes 4--35~km (20--400~Pa for a surface pressure of 610~Pa). 
Despite having selected about a third of the spectral information in the CO$_2$ band, we note that redundancy in functional derivatives peaking at similar levels is still large.
On the other hand, information on the temperature at altitudes 35--55~km (2--20~Pa) exclusively comes from the two spectral points at the core of the CO$_2$ band, at wavenumbers 667.7 and 668.4~cm$^{-1}$. We also notice that these two functional derivatives have broader full width at half maximum (being about 15~km) compared to the contribution functions peaking at 4--35~km, which have half-widths of $\sim$10~km.
This gives a qualitative estimate of the vertical resolution of the retrieved profile, which varies between 1 scale height (in the lower troposphere) and 1.5 scale heights (in the middle troposphere).

\subsubsection{Building the first guess temperature profile}

In a Bayesian approach, one possibility would be to compute both the \textit{a priori} temperature profile and its covariance matrix $\mathbf{S_a}$  from a climatology of temperature profiles, such as provided by the Mars Climate Database. The first guess can also be different from the \textit{a priori} temperature profile.
This is the methodology employed for instance by \citeA{Grassi2005} for PFS data analysis.
However, this method has its limits when atmospheric conditions depart significantly from the "climate" scenario of the MCD \cite<in the event of a global dust event, for instance , see >[]{Wolkenberg2018}. 
We choose instead the same approach as \citeA{Conrath2000}: for each observation, we build the first guess from  TIRVIM spectra themselves. Furthermore, this first guess is chosen as our \textit{a priori} profile. Hence, we are not here in a purely Bayesian formalism, as our \textit{a priori} profiles already contain information from the spectra. 
The method of \citeA{Conrath2000} to build this first guess consists of computing a set of  brightness temperatures from the radiance at several wavenumbers within the CO$_2$ band (in our case, we use the set of 50 wavenumbers described above), corrected for atmospheric transmission and surface contribution. This set of brightness temperatures is then attributed to different atmospheric pressure levels based on the contribution functions \cite<see equations 13 and 14 of>[]{Conrath2000} to build a first rough temperature profile, typically constrained between 10 and 40 km. This profile is then extrapolated upwards and downwards, using a method that differs from  \citeA{Conrath2000}.
For the upper part, we force the profile to smoothly go back to a climatology based on four martian years (MY29--MY33) of diurnal averages of MCS temperature profiles, binned every 5° of Ls and 5° of latitude, zonally averaged.
Regarding the extrapolation downwards, after several trials, we choose to impose the slope of the temperature profile in the first scale height based on an empirical relationship (determined from simulated synthetic spectra) between the temperature vertical gradient in the first scale height and the difference in brightness temperature between 690 and 702~cm$^{-1}$. 
Finally, the entire profile is smoothed vertically by convolution with a Gaussian function with a 0.75 scale height half width.
Regarding the covariance matrix $\mathbf{S_a}$, it is defined like in \citeA{Conrath2000} as to filter non-realistic vertical oscillations:

\begin{equation}
    S_\mathrm{a_{ij}} = \exp{\left[- \frac{log(p(i)/p(j))^2} {2 c^2}\right]}
\end{equation}

with $c$ a correlation length chosen nominally as 0.75 scale heights. Like $\beta$, this parameter $c$ is read in an input file.

\subsubsection{Surface temperature and aerosol retrieval scheme \label{sec:inv:aero}}

We now focus on the range 780--1300~cm$^{-1}$ of the spectra.
The general idea is to iteratively update first guess values of the surface temperature $T_\mathrm{surf}$ and of the dust and water ice vertical mixing ratio profiles $q_\mathrm{d}$  and $q_\mathrm{i}$ by exploiting the calculated functional derivatives of the brightness temperature to these parameters (noted $K_\mathrm{Ts}$, $K_\mathrm{q_d}$ and $K_\mathrm{q_i}$). 
We emphasize that nadir spectra are generally not (or very weakly) sensitive to the vertical distribution of dust and water ice clouds, hence, we only retrieve a scaling factor to \textit{a priori} profiles of their mixing ratio.
The functional derivatives $K_\mathrm{q_d}$ and $K_\mathrm{q_i}$ we use actually relate to the change in brightness temperature associated with a \textit{relative} change in $q_\mathrm{d}$  and $q_\mathrm{i}$, rather than a change in absolute values of $q_\mathrm{d}$  and $q_\mathrm{i}$.
Minimization of a cost function leads to the following iterative increment in $T_\mathrm{surf}$, $q_\mathrm{d}$  and $q_\mathrm{i}$ in the case of the simultaneous retrieval of all three quantities:

\begin{subequations}
\begin{align}    
    T_\mathrm{surf (n+1)} = T_\mathrm{surf (n)} + \sigma_\mathrm{Ts}K_\mathrm{Ts}^T Y  \Delta I \label{eq:ts}\\
    q_\mathrm{d (n+1)} = q_\mathrm{d (n)}  (1 + \sigma_\mathrm{q_d}K_\mathrm{q_d}^T Y  \Delta I ) \label{eq:taud}\\
    q_\mathrm{i (n+1)} = q_\mathrm{i (n)}  (1 + \sigma_\mathrm{q_i}K_\mathrm{q_i}^T Y  \Delta I ) \label{eq:taui}
\end{align}
\end{subequations}

\begin{equation}
\label{eq:Y}
    Y= \left( \sigma_\mathrm{Ts} K_\mathrm{Ts}K_\mathrm{Ts}^T+
    \sigma_\mathrm{d} K_\mathrm{q_d}K_\mathrm{q_d}^T+
    \sigma_\mathrm{i} K_\mathrm{q_i}K_\mathrm{q_i}^T+\frac{1}{\gamma}S_e \right)^{-1}
\end{equation}

where $\Delta I$ is the difference between the synthetic spectrum -- computed with surface temperature $T_\mathrm{surf (n)}$, aerosol mixing ratio profiles $q_\mathrm{d (n)}$ and $q_\mathrm{i (n)}$ -- and the TIRVIM spectrum in the range 780--1300~cm$^{-1}$; $\sigma_\mathrm{Ts}$, $\sigma_\mathrm{d}$ and $\sigma_\mathrm{i}$ are the \textit{a priori} covariance on surface temperature (set to 8K), dust and water ice cloud opacity (both set to a factor of three); and $\gamma$ is a parameter used to assign more or less weight to the observations with respect to our \textit{a priori} knowledge (determined similarly as $\alpha$ in equation~\ref{eqT2}). 
As for the atmospheric temperature retrieval, we do not use all spectral points in the retrieval of these quantities, especially since dust and water ice exhibit smooth spectral features. A set of 78 out of 805 spectral points in the range 785--1295~cm$^{-1}$ was carefully selected to optimize the information content and gain calculation time. This selection was based on a study of the Jacobians for surface temperature, dust and water ice opacities (derivatives of brightness temperature over these scalar quantities), computed for synthetic data. We retain more points near the peaking values of these Jacobians (one point every $\sim$3 cm-1). Away from these regions of maximum information content, our sampling is sparser (one point every $\sim$10 cm-1).

We emphasize that dust and water ice cloud opacities cannot be retrieved at all local times. 
This issue was already discussed e.g. by \citeA{Smith2019Icar} from the analysis of THEMIS broadband thermal infrared images, and also by \citeA{Wolkenberg2018} from PFS/Mars Express spectra.
The challenge in retrieving aerosol opacities from these nadir-viewing instruments is that sensitivity to dust and/or water ice (given here by $K_\mathrm{q_d}$ and $K_\mathrm{q_i}$) tends to be zero in the event when most of the dust (or water ice) mass load lies at altitudes where atmospheric temperature is close to surface temperature. 
In that case, no absorption nor emission band will be visible in the spectra, irrespective of the aerosol load, and we will not be able to determine the aerosols' opacity.
This situation is mostly encountered near 7--8~AM and 7--8~PM, as we will see in section~\ref{sec:results:dust}.
Of course, in the event when there are no aerosol absorption or emission bands seen in TIRVIM data, but at the same time the surface -- atmosphere temperature contrast is high, even a small amount of dust (or water ice) should have produced an absorption or emission feature, hence we would confidently conclude that the atmosphere is clear.

We have also added the option to not retrieve a given variable among surface temperature, dust and water ice opacity in  case a TIRVIM spectrum is too noisy and does not exhibit a clear dust or ice spectral feature.
Indeed, we noticed that in such situations, our retrieval algorithm sometimes fitted the data (or rather, fitted the noise) with a combination of values of surface temperature, dust or ice opacities that were unrealistic. 
Retrieving only a subset of parameters in this situation help regularize the inverse problem.
In more detail, we start by computing the standard deviation of TIRVIM brightness temperature in the range 1210--1290~cm$^{-1}$ as a proxy for the noise level (hereafter called noise$_\mathrm{1250}$). 
Indeed, beyond 1200~cm$^{-1}$, we find that the NER provided by the instrument team often underestimate the actual noise level so that we chose to estimate it from the data itself in this rather flat (in brightness temperature) spectral region.
We then compute the contrast in TIRVIM average brightness temperature between the range 1060--1130~cm$^{-1}$ and the range 1225--1290~cm$^{-1}$ (hereafter called contrast$_\mathrm{dust}$) and the one between 810--860~cm$^{-1}$ and 1225--1290~cm$^{-1}$  (hereafter called contrast$_\mathrm{ice}$).
Different configurations are considered:
\begin{itemize}
\item If spectra are noisy (arbitrary threshold set to noise$_\mathrm{1250}$ >6K) and contrast$_\mathrm{dust}$ and contrast$_\mathrm{ice}$ are smaller than noise$_\mathrm{1250}$: we retrieve surface temperature only;
		\item If spectra are noisy and contrast$_\mathrm{dust} <$ noise$_\mathrm{1250}$ and contrast$_\mathrm{ice} >$ noise$_\mathrm{1250}$: we simultaneously retrieve surface temperature and water ice opacity;
		\item If spectra are noisy and contrast$_\mathrm{dust} >$ noise$_\mathrm{1250}$ and contrast$_\mathrm{ice} <$ noise$_\mathrm{1250}$: we simultaneously retrieve surface temperature and dust opacity;
		\item Else, we simultaneously retrieve surface temperature, dust and water ice opacity.
	\end{itemize}
The retrieval of only one or two parameters among surface temperature, dust and water ice opacity follows the same form as equations  ~\ref{eq:ts}--\ref{eq:Y}. 

In order to evaluate the level of confidence we have on dust and water ice cloud retrievals, we adopt a similar approach as \citeA{Smith2019Icar} and define the 1-$\sigma$ relative error on dust opacity $dust_\mathrm{err}$ as the ratio between the 1-$\sigma$ noise level (in brightness temperature; computed from the instrument NER estimates) at 1100~cm$^{-1}$ and the functional derivative $K_\mathrm{q_d}$, computed at 1100~cm$^{-1}$ as well. 
When this ratio equals one, this means that a 100\% change in dust opacity results in a change in brightness temperature similar to the 1-$\sigma$ noise level.
In other words, the 1-$\sigma$ error on dust opacity would be 100\% in this case. We arbitrarily define a dust quality flag that  considers only dust retrievals with $dust_\mathrm{err}$ values lower than one. Similarly, for ice, we define $ice_\mathrm{err}$ as the ratio between the noise level and $K_\mathrm{q_i}$, both estimated at 820~cm$^{-1}$.
Examples of typical errors on dust and water ice cloud opacities will be detailed in section~\ref{sec:synthe}.

\subsubsection{Choice of \textit{a priori} values for aerosols and setting of the cloud condensation level}

Another challenge is related to the fact that even though the functional derivatives used in the retrieval in equation~\ref{eq:ts}--\ref{eq:taui} are updated at each iteration, they are themselves first computed based on prior information on the vertical distribution of dust and water ice clouds that can be erroneous and hamper our retrievals. 
For instance, if we \textit{a priori} assume that a cloud lies at an altitude where atmospheric temperature is warmer than the surface temperature, the functional derivative $K_\mathrm{q_i}$ will be positive in sign (the cloud is assumed to be seen in emission). 
However, if the cloud is actually at an altitude where atmospheric temperature is colder than the surface (i.e. the cloud is seen in absorption in TIRVIM data), the retrieval scheme will fail to reproduce the observed absorption band as we only retrieve a scaling factor to an initial mixing ratio profile. 
This actually implies that there are situations where we can say that TIRVIM spectra contain some level of information on the cloud altitude. This piece of information has to be exploited, if possible.
More generally speaking, the choice of \textit{a priori} values for the vertical distribution of $\mathrm{q_d}$ and $\mathrm{q_i}$ can strongly influence our results while in the case of atmospheric temperature retrieval, this choice is less critical.

While dust is rather ubiquitous, water ice clouds can be discrete and exhibit a greater spatio-temporal variability than dust, which makes their choice of \textit{a priori} properties not trivial. 
It would be tempting to run several retrievals for each observation, starting from different \textit{a priori} values for dust and ice opacities, assuming different cloud condensation levels, and then select \textit{a posteriori} the solution (or solutions) that best match the considered TIRVIM spectrum.
However, this would be too costly in computation time.
We thus choose an intermediate path: in a first stage of our algorithm, we explore various combinations of dust opacities ($\tau_\mathrm{dust}$=0.1, 0.3, 0.6), ice opacities ($\tau_\mathrm{ice}$=0.1, 0.3, 0.7, 1.5 and 4) and cloud altitudes (three different values, among which the condensation level derived from TES water vapor climatology).  
We exclude combinations that are not realistic, e.g. large values of $\tau_\mathrm{ice}$ at high altitudes.
For each combination, we compute a first guess synthetic spectrum in the range 800--1290~cm$^{-1}$.
To compute these spectra, the surface temperature is evaluated directly from the brightness temperature of TIRVIM spectra in the range 1240--1290~cm$^{-1}$, a portion of the spectrum that is rather transparent.
We pick up the combination of parameters that best match the considered TIRVIM spectrum at this stage as \textit{a priori} values, and then start a retrieval.
This way, we make sure to start our retrievals with prior values that match basic features of TIRVIM data (dust or ice seen in emission or absorption ; hints for a clear or a high dust load in the atmosphere, for instance) and hence, with sensible values of $K_\mathrm{q_d}$ and $K_\mathrm{q_i}$.

\subsection{Algorithm steps \label{sec:algo_step}}

The different steps of our algorithm, including those dealing with the special cases mentioned above, and the link between the retrieval of aerosol opacities, surface and atmospheric temperature, are summarized in this section.

First, a pre-processing step discards observations with known issues and extract ancillary data (surface pressure and CO$_2$ profile from the MCD, surface emissivity and climatological water vapor from TES, temperatures from MCS climatology, etc).
One NetCDF file per TGO orbit is created, which contains TIRVIM spectra with the aforementioned ancillary information. Typically, one file contains hundreds to thousands of spectra.
After this pre-processing step, the retrieval algorithm works as follows :

\begin{enumerate}
    \item Read in the NetCDF file and a list of optional parameters in a text-format input file, begin loop on all observations within this file.
	\item Assuming a small dust and water ice opacity (0.03), build first guess temperature profile from the spectrum and compute the condensation level $p_\mathrm{cond~TES}$ of water ice clouds, based on this first guess profile and TES water vapor climatology.
	\item Compute noise$_\mathrm{1250}$, contrast$_\mathrm{dust}$ and contrast$_\mathrm{ice}$.
	\item If both absolute values of contrast$_\mathrm{dust}$ and contrast$_\mathrm{ice}$ are smaller than noise$_\mathrm{1250}$, go to step 8 (the spectrum is considered $\sim$flat and there is no need to explore a range of first guess values), else continue.
    \item Build a family of possible and realistic combinations of  $\tau_\mathrm{dust}$, $\tau_\mathrm{ice}$, $p_\mathrm{cond}$ (among which the condensation level $p_\mathrm{cond~TES}$ determined above, but also encompassing potential lower and/or upper cloud, depending on the first value $p_\mathrm{cond~TES}$).
    \item For each of the above combinations: build the corresponding first guess temperature profile from TIRVIM spectrum in the range 660--740~cm$^{-1}$; estimate surface temperature from TIRVIM spectrum in the range 1240--1290~cm$^{-1}$; compute a first synthetic spectrum; compare with the TIRVIM spectrum and store the corresponding $\chi^2$. 
    \item Select the combination of first guess values for $\left( \tau_\mathrm{dust}, \tau_\mathrm{ice}, p_\mathrm{cond} \right)$ corresponding to the minimum $\chi^2$ ; keep the corresponding surface and atmospheric temperatures as first guesses as well.
    \item Compute a first synthetic spectrum with the first guess values, this time along with the computation of the functional derivatives for all quantities. Initialize $\chi^2$.
    \item Start a retrieval loop. Different configurations are considered.
    \begin{itemize}
        \item If spectra are noisy and contrast$_\mathrm{dust}$ and contrast$_\mathrm{ice}$ are smaller than noise$_\mathrm{1250}$: make one step of surface temperature retrieval only;
		
		\item If spectra are noisy and contrast$_\mathrm{dust} <$ noise$_\mathrm{1250}$ and contrast$_\mathrm{ice} >$ noise$_\mathrm{1250}$: make one  step of simultaneous surface temperature and water ice opacity retrieval;
		\item If spectra are noisy and contrast$_\mathrm{dust} >$ noise$_\mathrm{1250}$ and contrast$_\mathrm{ice} <$ noise$_\mathrm{1250}$: make one  step of simultaneous surface temperature and dust opacity retrieval;
		\item Else: make one step of simultaneous surface temperature, dust and water ice opacity retrieval following eq.~\ref{eq:ts}--\ref{eq:taui}.
	\end{itemize}
	
	\item Re-compute a synthetic spectrum with updated quantities (surface temperature and / or $\tau_\mathrm{dust}$ and / or $\tau_\mathrm{ice}$), calculate the functional derivatives for atmospheric temperature.
	\item Make one step of atmospheric temperature retrieval following eq.~\ref{eqT1}.
	\item Re-compute a synthetic spectrum with updated atmospheric temperature, then update $\chi^2$: if the change in $\chi^2$ compared to the previous value is less than 2\%, or if the number of iteration is $>$ 9, then: end the retrieval loop, write outputs and go on with the next observation; else go back to step 9.
		
\end{enumerate}

A solution is reached most of the time in 4 or 5 iterations. 
Even though only a fraction of the TIRVIM spectrum is exploited in equations~\ref{eqT1},~\ref{eq:ts}--\ref{eq:taui} and in $\chi^2$ calculations, we do compute a synthetic spectrum on the full spectral range of TIRVIM after convergence, that we keep for quality checks.

Our algorithm is not Bayesian, in particular as the so called \textit{a priori} temperature profile  already contains information from the data themselves. We note that it is also the case for the reference algorithm used for MGS/TES data, developed by \citeA{Conrath2000}.
Rapid convergence is achieved mostly because we start the retrievals from first guess values already close to the "true" atmospheric  state; however, in the next section, we will also show examples starting from different \textit{a priori} profiles to demonstrate that our code still performs well in these conditions. 


\section{Synthetic retrievals and error analysis\label{sec:synthe}}

\subsection{Synthetic observations and approach}

The precise tuning of the retrieval algorithm described previously results from several phases of trials and development, depending on how well the algorithm performs against synthetic measurements.
To build these synthetic observations, we extract surface pressure, surface and atmospheric temperature as well as aerosol mixing ratio profiles from the Mars Climate Database for the scenario corresponding to Martian Year 25. We choose this year as it features a global dust event at Ls 190--240° \cite{Smith2002}.
We extract these data at longitude 0° for various conditions: at five seasons (Ls= 0, 90, 180, 210 and 270°), 8 local times (every three hours) and 11 latitudes (from 75S to 75N, every 15°), hence 440 spectra in total. 
We then generate synthetic TIRVIM spectra with our forward model, then add noise (with realistic spectral dependency of the noise), and run our retrieval algorithm.
This exercise is often referred to in the literature as an Observing System Simulation Experiment (OSSE).

Generally speaking, this exercise allows to test the sensitivity to, for instance, assuming a different vertical profile of dust and water ice clouds in the retrieval set-up compared to the forward model used to generate the synthetic data, and its impact on the retrieved quantities.
It is also useful to identify flaws in our retrieval pipeline, fine-tune the sampling in wavenumber and refine our error analysis.
We do not test the sensitivity of the retrieved quantities to an error in surface pressure, surface emissivity, spectroscopic database or aerosol particle sizes: these parameters are the same in the forward model and retrieval pipeline. 
Hence, the only forward model error that we consider comes from the different vertical mixing ratio profiles assumed for dust and water ice. 
Despite this quite favorable setting, we will see that many challenges arose solely due to the intrinsic degeneracy of the inverse problem and/or noise in the spectra.

\subsection{Robustness of the synthetic retrievals: aerosols}

\begin{figure}
  \centering
  \includegraphics[width=0.9\linewidth]{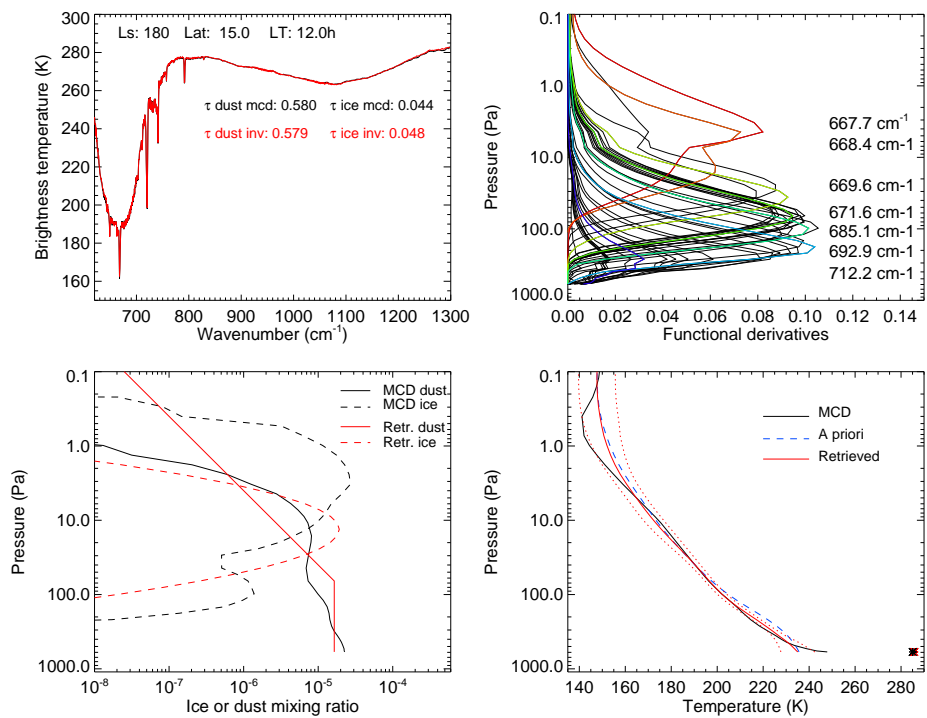}
  \caption{Summary of a synthetic retrieval for a MCD scenario extracted at L$_s$=180°, latitude 15°N, local time 12~AM (noon). Top, left: Synthetic TIRVIM spectrum (black) along with the best fit (red). Retrieved and MCD dust and water ice cloud integrated opacities are indicated. Top, right: functional derivatives of the temperature as a function of pressure, for 50 different wavenumbers within the CO$_2$ band. Several wavenumbers are highlighted in different colors and labeled. Bottom, left: Mixing ratio vertical profiles for dust and water ice as taken from the MCD (used to generate the synthetic spectrum, in black) and as derived from the retrieval process (in red ; note that only a scaling factor to a generic \textit{a priori} profile is retrieved). Bottom, right: Temperature vertical profile from the MCD (used to generate the synthetic spectrum, in black), \textit{a priori} profile built from the synthetic spectrum (dashed blue line), and retrieved profile (red), with error bars in red dotted lines. The black and red stars stand for the MCD and retrieved surface temperature, respectively.}
  \label{fig:synthe1}
\end{figure}

Overall, synthetic retrievals perform quite as expected: cases with a warm surface temperature (hence high SNR) yield the most robust results for all retrieved parameters. For such cases, the integrated opacities of dust and water ice clouds are well estimated even if the assumed simplified vertical distributions of aerosols differs from the "actual" one (taken from the MCD), thanks to the large surface-atmosphere temperature contrast. A typical example of such a favorable case, with a surface temperature of 285K, is presented in Figure~\ref{fig:synthe1}.

For surface temperatures lower than 230K, the fraction of retrievals that fail to correctly retrieve dust and water ice opacity increases.
These unsatisfactory cases fall in two categories: either the quality criterion for dust or ice retrieval (defined in section~\ref{sec:inv:aero}) is not met, or this criterion is met (in the sense that we consider there is enough sensitivity to ice or dust in the spectra to trust the results) but the retrieved integrated ice or dust opacity differs significantly from the "true" (MCD) one nonetheless.
This occurs (but not systematically) when the assumed dust mixing ratio at altitudes where the temperature is similar to surface temperature differs significantly from that of the MCD (in other words, TIRVIM spectra are blind to dust loading at some pressure levels but as we integrate the whole dust column opacity, an error in this blind zone impacts the integrated optical depth) ; and/or when there is a high level of degeneracy between surface temperature and dust and/or ice, in particular when spectra exhibit shallow dust and/or ice features.

\begin{figure}
  \centering
  \includegraphics[width=0.6\linewidth]{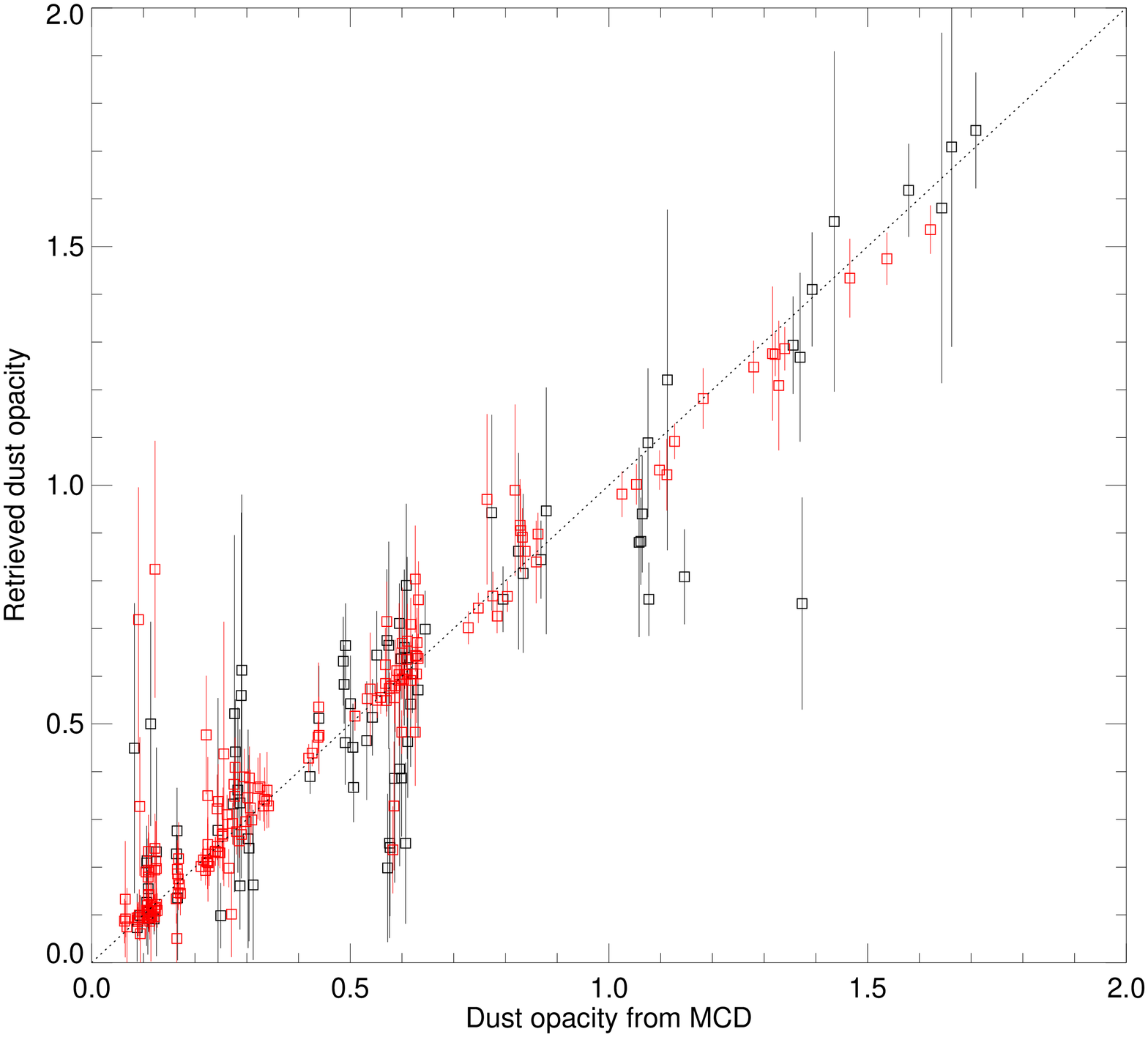}
  \caption{Retrieved versus input (MCD) integrated dust opacity in our OSSE. Only cases that passed the dust quality flag are shown (all squares, back and red). In red are cases for which the temperature constrast between surface and lower atmosphere (at 4~km) is greater than 20K.
  The dashed line highlights a one-to-one correspondence.}
  \label{fig:correlation_dust}
\end{figure}

About 60\% of the retrievals pass our quality filter for dust.
We present a summary of all these good quality-flagged retrieved dust opacities in Figure~\ref{fig:correlation_dust}. 
Most of the scenes that pass this quality filter also exhibit large ($>$20K) surface--lower atmosphere temperature contrast, which is consistent as this situation corresponds to a greater sensitivity to dust.
Our algorithm performs well even in dusty conditions, ie. with an integrated optical depth greater than one. However, this is probably too optimistic, as we neglect multiple scattering by dust in both the forward model used to generate the synthetic spectra and the retrieval pipeline. 
The Pearson correlation coefficient between MCD and retrieved dust opacity is 0.94. The standard deviation of the difference between retrieved and "true" dust opacity is 0.12, which is larger than the mean absolute 1-$\sigma$ error in dust estimated from the ratio between noise level and $K_\mathrm{d}$, which yields a value of 0.08. This suggests that our formal error is underestimated, but this is not surprising, given different sources of potential biases.
We find that 12\% of the dust retrievals that do pass the dust quality filter suffer from a significant bias (compared to the "true" dust opacity), ie. two times larger than our 1-$\sigma$ error estimate.
Some outliers are indeed visible in Figure~\ref{fig:correlation_dust}, and we also identify several groups of cases where dust is systematically over or underestimated in the retrievals.
The physical reason behind these outliers will be further discussed below for a few representative cases.

\begin{figure}
  \centering
  \includegraphics[width=0.9\linewidth]{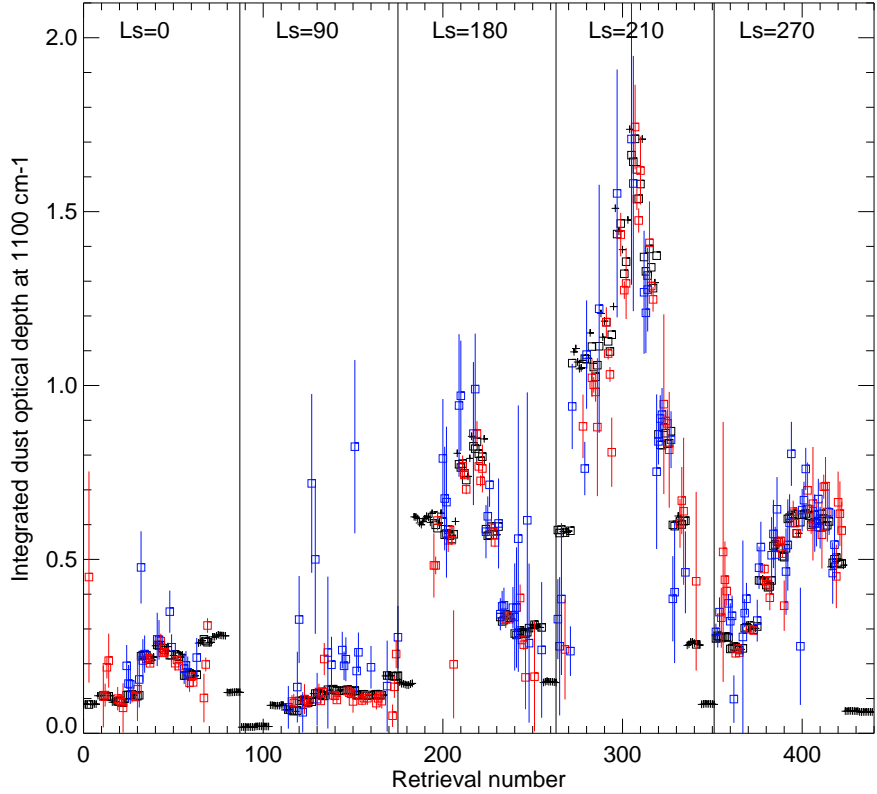}
  \caption{Evaluation of the robustness of dust retrievals sorted by season.  Data points are grouped by values of Ls. Within each group, data are sorted by latitude from 75S to 75N, more clearly seen in Figure~\ref{fig:dust_mcd_270}. Black crosses represent integrated dust opacities from the MCD, used to generate synthetic spectra in the OSSE. Blue squares are retrieved dust opacity that passed the dust quality flag for local times 0AM, 3AM, 6AM and 9PM, red squares for local times 9AM, 12AM, 3PM and 6PM. 
  Vertical bars show the $\pm$ 1-$\sigma$ error as defined in section~\ref{sec:inv:aero}. 
  When no colored square is associated to a black cross, it means that the retrieval did not pass the dust quality flag, hence they were not plotted.}
  \label{fig:dust_mcd_all}
\end{figure}

Retrieved and MCD dust opacities are shown for different seasons, latitudes and local times in Figure~\ref{fig:dust_mcd_all}. A close-up of the Ls=270° case is shown in Figure~\ref{fig:dust_mcd_270}.
Daytime retrievals (red squares with error bars in Figures~\ref{fig:dust_mcd_all} and \ref{fig:dust_mcd_270}) are in general more robust, with smaller associated retrieval errors. This is mostly explained by a large surface-atmosphere temperature contrast and warmer temperatures (hence higher S/N ratio).
Systematic overestimation of dust is visible in particular at Ls=90° at low latitudes (15S--30°N) for nighttime retrievals, where an integrated dust opacity of $\sim$0.2 is retrieved at night, instead of $\sim$0.1 in the MCD. Daytime retrievals on the other hand are robust at this season and latitudes. These different behaviors create a spurious daily cycle of dust in the retrievals, not present in the MCD.
Two examples of overestimated dust are shown in Figures~\ref{fig:synthe5} and ~\ref{fig:synthe6}. In the first case (latitude 15N, midnight), an optically thick cloud centered at $\sim$100~Pa prevents the correct retrieval of atmospheric temperature in the lower atmosphere (see the functional derivatives of temperature in Figure~\ref{fig:synthe5}). 
The retrieved surface-atmosphere temperature contrast is underestimated, and as a result, the dust load is overestimated by our algorithm. Indeed, there is a degeneracy in the solution of the inverse problem between dust opacity and temperature contrast, and we see here an example of a compensation of two errors.
In the second example in Figure~\ref{fig:synthe6} (latitude 30N, 3AM), the atmospheric temperature is well estimated. Here, the issue seems to be the \textit{a priori} dust vertical distribution, assumed well-mixed in the first two scale heights while it quickly decreases with height in the MCD profile. Because atmospheric temperature at 50--200~Pa is close to the surface temperature, we are not sensitive to the dust in this altitude range, but it will be anyhow added up in the calculation of integrated optical depth.
Hence, while the amount of dust in the lower part of the atmosphere seems well estimated, the retrieved total integrated dust opacity is mechanically overestimated.

We detail another case study of challenging dust retrieval, this time during the global dust event of MY25 (Ls=210°) and shown in Figure~\ref{fig:synthe3}. Here, in spite of the high dust opacity (1.37) in the MCD scenario, the synthetic spectrum exhibits a very shallow dust emission feature. This is due to the modest surface-atmosphere temperature contrast in the first scale height, combined to  the presence of dust at high (and colder) altitudes in the MCD profile at pressures lower than 60~Pa: the shallow dust feature results from a combination of dust thermal emission at various temperatures.
Here, the underestimation of retrieved dust opacity by a factor of two by our algorithm is partly due to the wrong assumption on dust vertical distribution above 60~Pa, and partly due to the fact that the large dust opacity increases the degeneracy between surface temperature and dust retrieval. 
As a consequence, here, a slightly warmer surface temperature and less dust in the lower atmosphere yields a similar brightness temperature as the MCD scenario, that has a slightly colder surface and higher dust loading but over a greater column, with different emission temperatures. These three examples illustrate well some of the subtle challenges encountered (that we will have to keep in mind when interpreting actual TIRVIM data). In spite of this, we recall that overall, nearly 90\% of the dust retrievals that pass our quality flag are found satisfactory.

\begin{figure}
  \centering
  \includegraphics[width=0.7\linewidth]{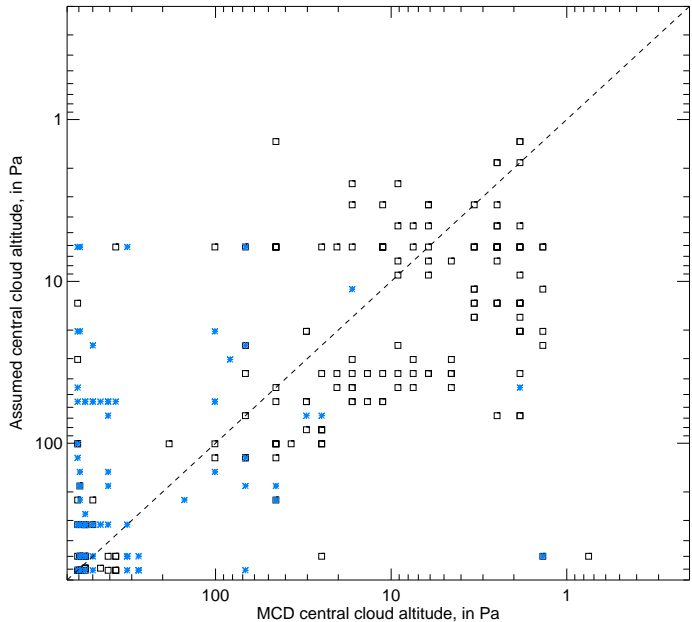}
  \caption{Cloud altitude as set up in the retrieval algorithm (see  section~\ref{sec:algo_step}) versus MCD central cloud altitude, in Pa. These values have been normalized to 610~Pa. Only cases that pass our ice quality flag are included and shown as black squares and blue stars. Blue stars highlight cases where ice opacity retrievals were significantly biased with respect to MCD values, despite having been flagged as robust ice retrievals.}
  \label{fig:cloud_alti}
\end{figure}

Regarding cloud retrievals, about 62\% of the retrievals pass our water ice quality filter.
However, 32\% of these cases are significantly biased compared to the MCD values: this fraction is almost three times greater than that for dust retrievals.
Furthermore, we obtain a correlation of 0.74 between MCD and retrieved ice integrated optical depth. 
This confirms the overall worse performance of ice retrievals compared to dust retrievals.
We first investigate whether our assumed cloud altitude is realistic or not.
Figure~\ref{fig:cloud_alti} shows the MCD cloud "altitude" (pressure level of maximum ice mixing ratio) versus the cloud altitude set during the retrieval. 
Overall, the correspondence between the two is rather good, except for low altitude clouds (in the MCD) whose altitude is not well captured. Actually, most of the biased cloud retrievals correspond to low altitude clouds in the MCD (see Figure~\ref{fig:cloud_alti}), with large opacities ($>$3) in the MCD scenario. In these situations, the retrieval of surface temperature and/or lower atmospheric temperature is challenging, which in turn impacts the cloud opacity retrieval.

We also note that even a moderate error in the assumed cloud altitude can significantly impact the retrieved optical depth.
An example of a wrong determination of water ice cloud opacity linked to an error in the assumed water ice vertical mixing ratio profile is shown in Figure~\ref{fig:synthe2}.
Here, the MCD profile features a vertically extended cloud, from 3 to 300~Pa, with an integrated opacity of 0.325, while the retrieval assumes a thinner cloud centered at 10~Pa and yields a much lower ice opacity (0.136).
Still, the fit to the synthetic spectrum is satisfactory. 
This illustrates well a degeneracy frequently observed, typically in the surface temperature range 180--220K, between the ice opacity and the surface-atmosphere (where the cloud resides) temperature contrast. An optically thicker cloud located at an atmospheric temperature closer to the surface temperature yields a similar spectral signature as a less opaque cloud located at significantly colder (or warmer, if seen in emission) atmospheric temperature, compared to the surface.
The chosen example in Figure~\ref{fig:synthe2} is actually more subtle, as part of the vertically extended cloud is located in a region of  similar atmospheric temperature as the surface, making it invisible to retrievals.

Hence, as for dust retrievals, several examples of over- or underestimation of cloud opacity can be found while still presenting a good fit to the spectra, either linked to a wrong assumption on the ice vertical distribution and/or a degeneracy with surface temperature determination and/or wrong determination of atmospheric temperature near the surface, or a subtle combination of these effects.
A few difficult cases of combined degeneracy between surface temperature and both dust and ice opacity are also found.

\begin{figure}
  \centering
  \includegraphics[width=0.8\linewidth]{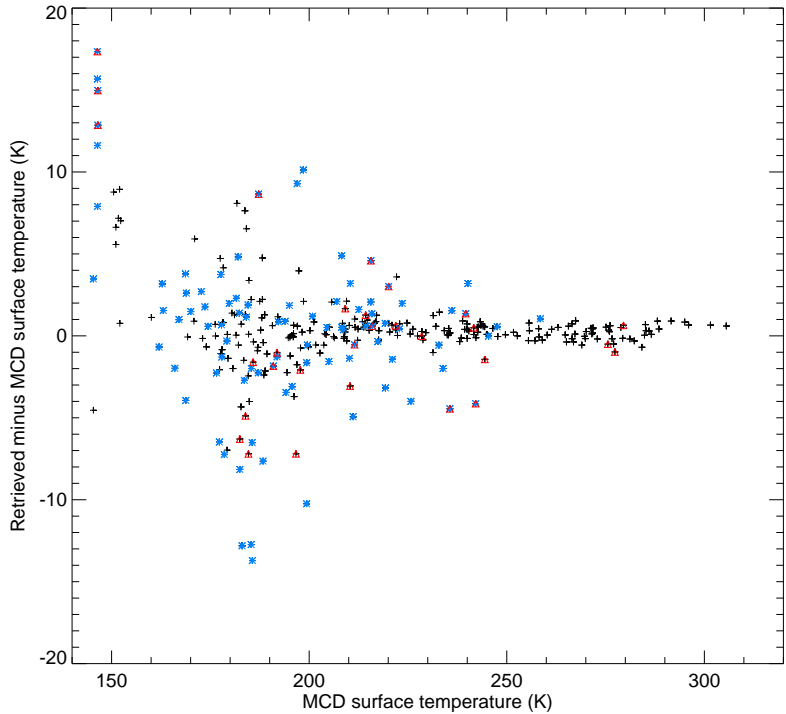}
  \caption{Error in surface temperature (defined as the difference between retrieved and "true" MCD surface temperature) as a function of the MCD surface temperature. We only show data points that pass either or both aerosol quality flags. We highlight as red triangles cases where the dust retrievals were significantly biased compared to MCD dust opacity, and as blue stars cases where ice retrievals were significantly biased (greater than two times the 1-$\sigma$ error). All black crosses correspond to the remaining cases, ie. that pass either one or both our aerosol quality flags and whose dust and ice opacity are not significantly biased compared to MCD values.}
  \label{fig:err_tsurf}
\end{figure}

The degeneracy between surface temperature and aerosol retrievals is further illustrated in Figure~\ref{fig:err_tsurf}, which displays the error in surface temperature (the a posteriori difference between retrieved and MCD surface temperature), only for cases that pass the dust or ice quality flags. 
Cases for which dust or water ice opacities are significantly biased with respect to the MCD values (defined by a difference with the MCD opacity greater than two times our estimated 1-$\sigma$ error) predominantly exhibit a greater error in surface temperature as well.
This illustrates the degeneracy between these quantities in the inverse problem: these errors compensate and a combination of erroneous values of surface temperature, dust and/or ice opacity can yield good fits to the synthetic TIRVIM spectra nonetheless. 
On the positive side, as mentioned at the beginning of this section, we note that for warm surface temperatures, the retrieval of all quantities remains robust.

\subsection{Robustness of the synthetic retrievals: atmospheric temperature \label{subsec:synthe_temp}}

Retrievals of atmospheric temperature profiles are found reliable in a vast majority of cases, even when dust, water ice or surface temperature retrievals are not robust.
An exception is for observations featuring low altitude opaque clouds. In this case, we can lose sensitivity to atmospheric temperature in the first scale height, as was shown in Figure~\ref{fig:synthe5} during the aphelion cloud belt, or in Figure~\ref{fig:synthe4} for a high latitude winter case.  
In such a situation, the retrievals are more impacted by a wrong \textit{a priori} near-surface temperature, which implies a wrong ice emission temperature and in turn, impacts the retrieved ice optical depth. 
The case in Figure~\ref{fig:synthe4} illustrates another challenge for nadir observations that constitutes the presence of a very steep temperature gradient in the lower atmosphere: here, the temperature increases from 163K to 211K between 490~Pa and 230~Pa, which represents a +48K increase over 7~km. Given the coarse vertical sensitivity of TIRVIM retrievals, such a steep gradient cannot be captured properly.
We note, however, that this issue is localized and that the retrieved atmospheric temperature behaves well above 10~km as it does depart from the \textit{a priori} profile and satisfactorily reproduces the local temperature maximum seen in the MCD profile at 100--200~Pa, where information content is high enough.

\begin{figure}
  \centering
  \includegraphics[width=0.9\linewidth]{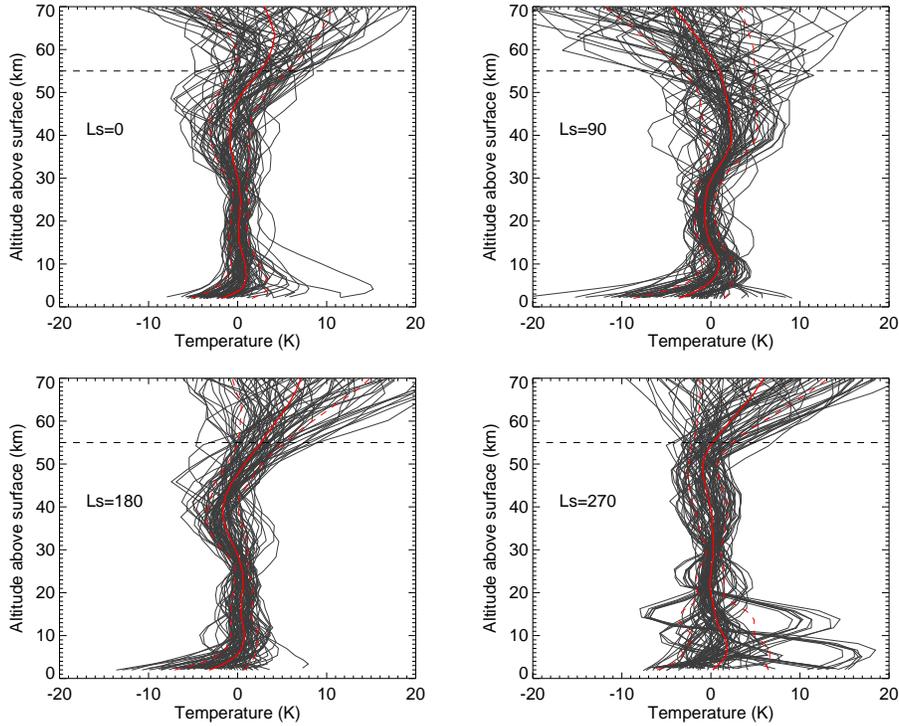}
  \caption{Vertical profiles of the difference between retrieved and MCD temperature profiles, sorted by seasons, as labeled. In each panel, the red solid line is the mean of the difference, and dashed red lines are 1-$\sigma$ standard deviation added and subtracted from the mean. The case Ls=210° is not shown but is almost identical to the case Ls=180°. Most of the error results from the rather coarse vertical resolution of TIRVIM (see Figure~\ref{fig:synthe_stat_temperature_AK} for comparison).}
  \label{fig:synthe_stat_temperature}
\end{figure}

Apart from the caveat described above, we find that the atmospheric temperature retrievals perform well. The difference between the retrieved temperature profiles and that from the MCD, MY25 (used to generate the synthetic observations), along with the mean and 1-$\sigma$ standard deviation of this difference, are shown in Figure~\ref{fig:synthe_stat_temperature}.
Most of the differences can be explained by the rather coarse vertical resolution of TIRVIM, as will be demonstrated in the following section.
In the region of maximum sensitivity (5--35~km), the mean of the difference is close to zero and the 1-$\sigma$ standard deviation generally lower than 3K.
In the 35--55~km (2--20~Pa) region, the vertical resolution is coarser. As a result, the retrieved temperature averaged over 35--55~km may be correct, although the slope may not (as is the case for Ls=180° in our  OSSE). 
This can result in rather large errors at a given pressure level in the range 35--55~km while the fit to the data is correct: this behavior is consistent with the information content of the data and illustrates the degeneracy due to the coarse vertical resolution. Above the altitude of maximum sensitivity (peaking near 2--3~Pa, or 50~km), the temperature profile smoothly returns to the \textit{a priori} profile, and results should not be interpreted. 
This will have to be kept in mind when  actual TIRVIM retrievals are discussed.
We report a similar issue in the first atmospheric scale height, in particular when very steep temperature gradient (in the MCD) are not caught by our retrieval, as mentioned previously and illustrated in Figure~\ref{fig:synthe4}. This is frequently associated to elevated water ice content near the surface and mostly occurs at Ls=270° in the winter high latitudes.
Finally, we note that the atmospheric temperature retrieved for global dust storm conditions is very satisfactory and is very similar to the Ls=180° case.

It might seem that our algorithm for atmospheric temperature retrieval performs well mostly because our \textit{a priori} temperatures profiles are well chosen, but it's not fully the case. To demonstrate the limited influence of the \textit{a priori} on our results, we have also run the 440 test cases of our OSSE with \textit{a priori} profiles $\pm$10K warmer or colder than our nominal profiles. Examples are shown in supplementary Figure~\ref{fig:supp_prior}.
The retrieved profiles converge towards a very similar solution  between a few kilometers above the surface and $\sim$3~Pa, independently of the choice of the \textit{a priori} profiles. Above 3~Pa, the profiles starts to diverge and go back to their own \textit{a priori} profile, consistently with the information content of the data.
The quality of the fit in the CO$_2$ band is overall a bit better when using our nominal \textit{a priori} profile, which is mostly due to slightly better fits of the core of the CO$_2$ band. This illustrates well the difficulty for our retrieval algorithm to depart far enough from the prior at 1--3~Pa (where information content is low) solely to fit 1--2 spectral points near 667 cm$^{-1}$, and justifies our choice of \textit{a priori} profile to mitigate this challenge. Another advantage is a slightly faster convergence of the retrievals ($\sim$4 iterations for our nominal profile instead of $\sim$5 iterations for the other two profiles). 
As we will rarely interpret observations above the 2--3~Pa level though, we show here that our results are largely independent of the choice of the \textit{a priori} profile.

Finally, we can also briefly compare the performance of our OSSE with that of \citeA{Grassi2005} done for synthetic PFS/Mars Express retrievals.
As expected, the authors highlighted a difficulty of their retrievals to capture strong vertical thermal gradients in the lowest levels of the atmosphere.
Based on 288 representative simulated spectra (extracted from the MCD at different seasons, local times and latitude), they found that their algorithm performed very well between 5 and 25~km, with systematic errors near zero and random error of the order of 2--3K. However, their retrieval exhibited a systematic error in the retrieved temperature above 30~km that reached 4K at 50~km, while the random error increased from 5 to 8K between 30 and 50~km.
Random errors are thus very comparable to ours, while we do not report any systematic errors up to 40~km.
\citeA{Grassi2005} also emphasized that for altitudes 40--50~km, the contribution of the a priori profile became more important than the weight of the data in constraining the solution. This is consistent with our analysis and is not surprising, as PFS and TIRVIM have similar spectral resolution and hence similar information content.

\subsection{Effect of averaging kernels on temperature profiles}
\begin{figure}
  \centering
  \includegraphics[width=0.9\linewidth]{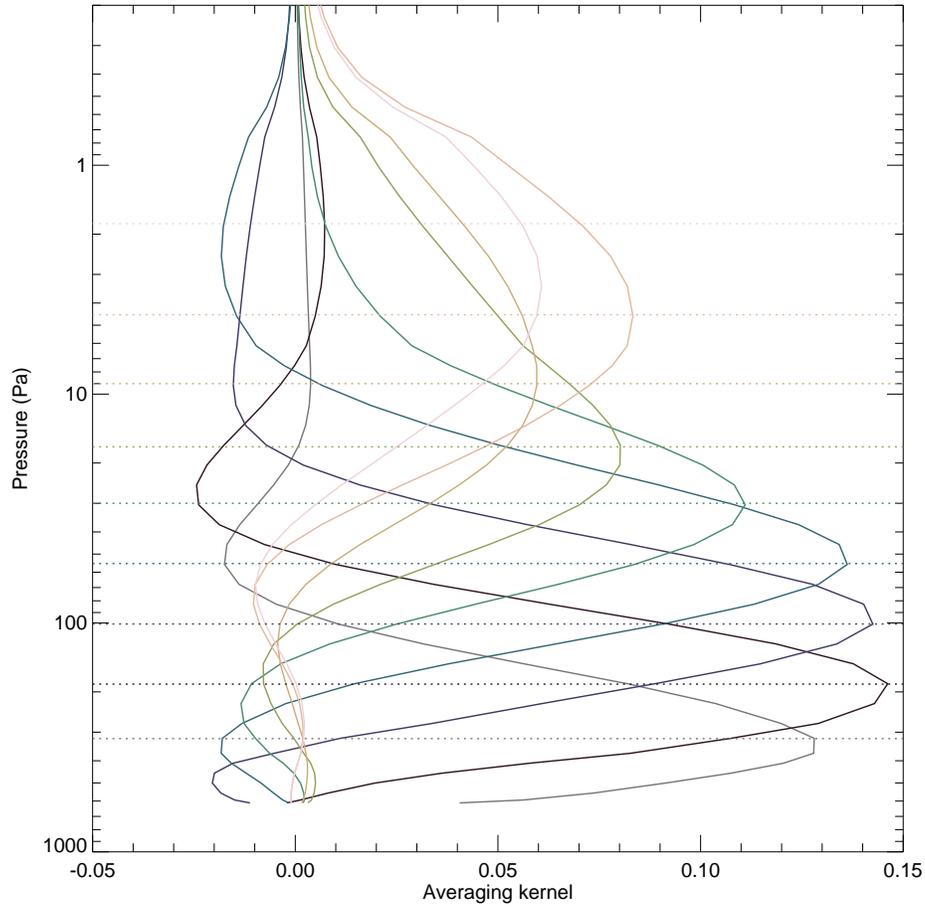}
  \caption{Example of nine rows of the averaging kernel matrix (solid lines), for nine pressure levels, materialized in dotted lines. This example was generated for atmospheric conditions extracted from the MCD for MY25, Ls=180°, latitude 15°N and local time 9~AM (same as Figure~\ref{fig:ex_kk}).}
  \label{fig:ex_AK}
\end{figure}

We can go one step further in the comparison of retrieved versus "true" profile by emulating the combined effects of TIRVIM coarse vertical resolution and the influence of the \textit{a priori} profile on the MCD profile.
Indeed, in theory, TIRVIM retrievals provide our best estimate $\mathbf{\hat{T}}$ of the true atmospheric state $\mathbf{T_\mathrm{true}}$, with the caveat that these retrieved profiles represent a smoothed version of the true state and partly contain \textit{a priori} information, in particular in altitude regions where information content is low. This is formally expressed by:

\begin{equation}
\label{eq_lissage}
    \mathbf{\hat{T}} = \mathbf{T_a} + \mathbf{A_k} (\mathbf{T_\mathrm{true}} -  \mathbf{T_a})
\end{equation}

with the averaging kernel matrix $\mathbf{A_k}$ defined by $\mathbf{A_k}= \mathbf{W K}$. 
This matrix is proportional to the weighting kernels and reflects the fraction of information coming from the \textit{a priori} and from the data. Examples of rows of $\mathbf{A_k}$ are presented in figure~\ref{fig:ex_AK}. The FWHM of a row gives an estimate of the vertical resolution of our retrievals and illustrates well how it is significantly larger (coarser) at higher altitudes. Note that for this example, rows of $\mathbf{A_k}$ at pressures higher than $\sim$4~Pa exhibit a sensitivity peaking at the corresponding pressure level, but that for lower pressures, the peak sensitivity occurs below that level (at higher pressures). In this example, our peak sensitivity does not reach higher than the 3~Pa level. 
More details on the use of averaging kernels can be found in \citeA{Rodgers2000}.

\begin{figure}
  \centering
  \includegraphics[width=0.9\linewidth]{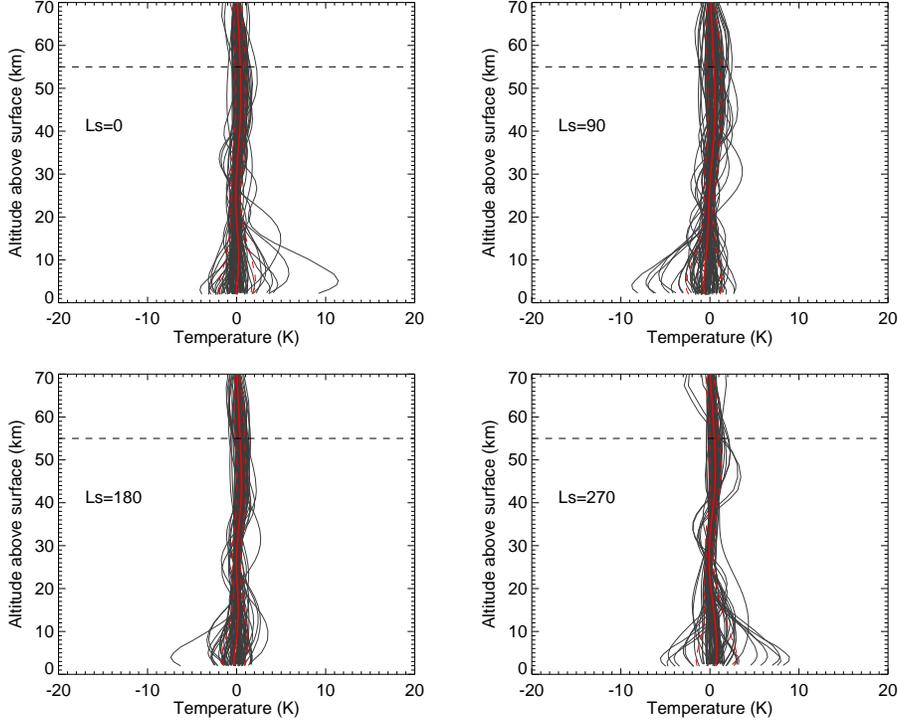}
  \caption{Same as Figure~\ref{fig:synthe_stat_temperature} but showing the difference between retrieved and "smoothed" MCD temperature profiles ($\hat{\mathbf{T}}$).}
  \label{fig:synthe_stat_temperature_AK}
\end{figure}

By applying equation~\ref{eq_lissage} to the "true" MCD profiles, we can thus run the MCD profiles through the retrieval filter, which degrades their vertical resolution and introduces a fraction of the \textit{a priori} temperature profile. 
If the retrieval algorithm works without flaw, the retrieved temperature profiles obtained in this OSSE should be almost identical (within the formal retrieval error) to the best estimates $\mathbf{\hat{T}}$ obtained with equation~\ref{eq_lissage}.
The updated comparison between our retrievals and $\mathbf{\hat{T}}$ profiles is shown in Figure~\ref{fig:synthe_stat_temperature_AK}.
As expected, for most cases, the difference is  smaller than 2K at all altitudes, since both profiles now go back to the same \textit{a priori}.
This signifies that 2K is a realistic estimate of the formal error linked to noise in the data and the non-linearity of the radiative transfer equation.
However, there are a few cases where the difference between profiles significantly exceeds error bars. This happens in particular at Ls=270°, for the challenging scenes with a thick low altitude cloud in the MCD. 
Here, the unexpected difference between the retrieval and $\mathbf{\hat{T}}$ stems from the wrong evaluation of the averaging kernel matrix $\mathbf{A_k}$. Indeed, our retrievals often fail to capture the actual high cloud optical depth, which impacts the calculated functional derivatives $\mathbf{K}$ and hence, $\mathbf{A_k}$ as well.
In the example shown in Figure~\ref{fig:synthe4}, the functional derivatives (wrongly) indicate that the sensitivity to the low atmospheric temperature should be rather good, which explains why $\mathbf{\hat{T}}$ is close to the MCD profile. However, the retrieved temperature is hampered here by a wrong estimate of water ice cloud opacity and the functional derivatives should not be as high in the first scale height.
We emphasize though that these situations are not frequent, and that overall, the retrieval algorithm behaves as expected.
To summarize, except for a few particular cases,  based on this OSSE, our confidence in the retrieved temperature profiles is quite high.

\section{Application to ACS/TIRVIM spectra and validation of the retrievals\label{sec:results}}
\begin{figure}
  \centering
  \includegraphics[width=0.9\linewidth]{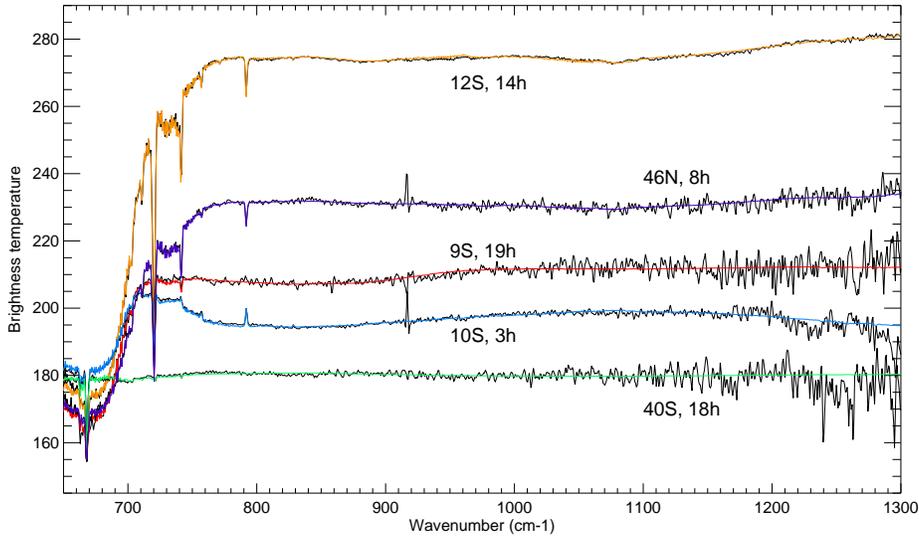}
\caption{Examples of TIRVIM spectra (in black) and best fits (in color). These five spectra were all acquired close to Ls=150° of Martian Year 34 and near 40°W. Latitudes and true solar local times are labeled; these cases were chosen to sample different surface and atmospheric temperatures.}

  \label{fig:ex_fits}
\end{figure}

We have applied our retrieval algorithm to the full TIRVIM dataset. The retrieved quantities, along with fits to the data, are distributed on the IPSL data center \cite{data}.
We illustrate examples of five TIRVIM spectra and their corresponding best fits at different latitudes and local times in Figure~\ref{fig:ex_fits}. 
In the following, we cross validate the atmospheric temperature profiles retrieved from TIRVIM against Mars Climate Sounder (MCS) profiles and continue with an evaluation of the retrieved dust integrated optical depth. We focus in this paper on the first few weeks of TIRVIM scientific operations.
Validation of water ice cloud opacity retrieval is deferred to future work for two reasons. Firstly, there were very few clouds at the season and latitude coverage corresponding to the first weeks of TIRVIM data (at Ls$\sim$150°). Secondly, the OSSE has shown that our cloud retrievals suffered from rather larger uncertainties and biases. We thus plan to improve our ice retrieval algorithm in the future before further validation and scientific exploitation (see discussion in section~\ref{sec:discuss}).

\subsection{Description of TIRVIM and MCS datasets and co-location method}

We focus on the validation of TIRVIM retrievals acquired during the first 45 sols of TGO science phase, ie. between March 13 and April 28, 2018. This corresponds to Martian Year 34 near the northern autumn equinox, from Ls=142° until Ls=167°. 
This dataset comprises both individual spectra (acquired during the first 6 days) and spectra resulting from the onboard averaging of 8 consecutive interferograms. 
Only 3\% of the retrievals did not pass the post-processing quality filters (mostly related to the quality of the fits, based on $\chi^2$), which is highly satisfactory. This is actually even more favorable than our exercise on synthetic retrievals (where overall, 9\% didn't pass our set of quality filters) and could be explained by the low number of challenging scenes (ie. elevated aerosol load and/or cold surface) at this season and for the latitudinal coverage considered here.

We choose to validate our results against Mars Climate Sounder (MCS) observations.
The MCS is a passive radiometer that records the thermal emission of the martian atmosphere in limb viewing geometry, in several thermal infrared channels, including three channels covering the 15-$\mu$m CO$_2$ band. The MCS team provides retrieved vertical profiles of the temperature with a vertical resolution of  5~km, from an altitude of typically $\sim$10--15~km (as low as 5~km in the event of a clear atmosphere) up to 80--90~km altitude, or from typically $\sim$300 to 0.02~Pa.
Due to the Sun-synchronous orbit of MRO, these profiles are only available near local times 3~AM and 3~PM, with a rather dense coverage in latitude and longitude.
The MCS retrieval algorithm includes a single-scattering approximation to account for scattering effects \cite{Kleinbohl2011} and accounts for  temperature and aerosol load variations along the line of sight \cite{Kleinbohl2017}.  
Temperatures derived from MCS have been used in many scientific studies \cite<e.g.,>[]{Lee2009,Kleinbohl2013}.  Because MCS samples a greater altitude range, achieves a higher vertical resolution, and its data are known for their high reliability, we choose to validate our TIRVIM temperature retrievals against MCS ones acquired at close locations, dates and local time.

\begin{figure}
  \centering
  \includegraphics[width=0.7\linewidth]{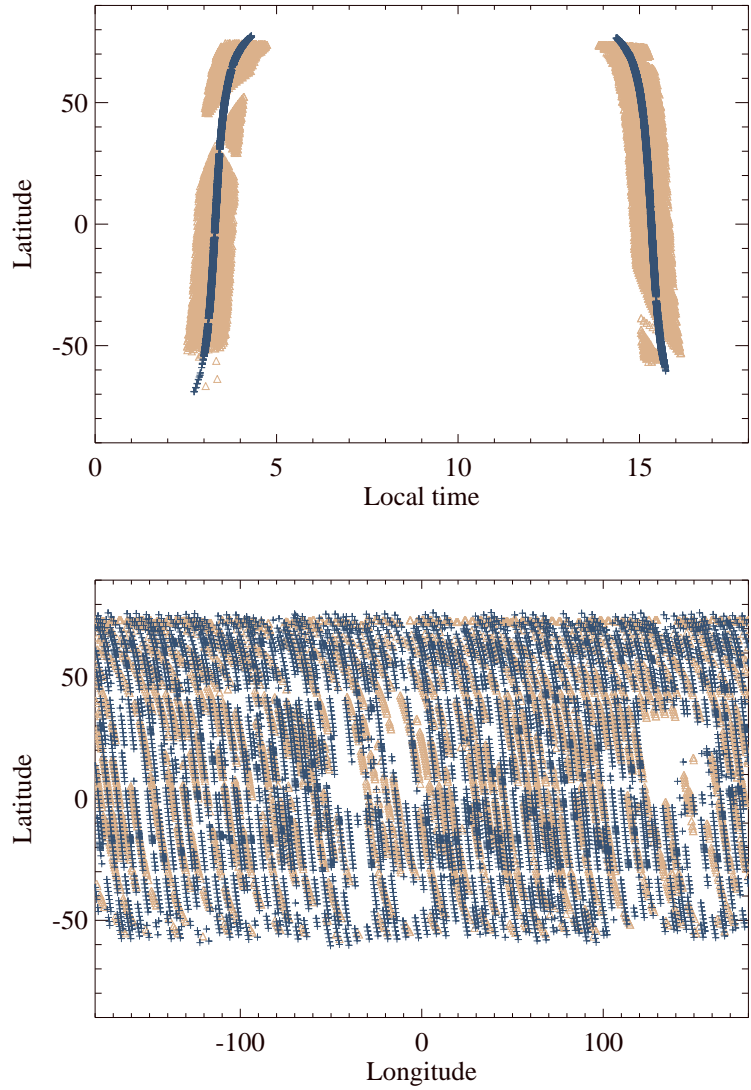}
\caption{Coverage of TIRVIM (orange triangles) and MCS (dark blue crosses) co-located observations over the course of the first 45 sols of TIRVIM science phase (Ls$\sim$150°). Top: latitude--local time coverage ; bottom: East longitude--latitude coverage for the near 3PM data only. Here we display existing MCS data co-located with TIRVIM, irrespective of whether the MCS profiles extend to near-surface altitudes or not.}

  \label{fig:coloc_coverage}
\end{figure}

For this exercise we select, for each TIRVIM observation, all MCS data acquired within 6$^{\circ}$ of longitude, 3$^{\circ}$ in latitude, and half an hour in local time of the considered TIRVIM profile (although we allow the date of observation to differ by $\pm$ one sol between the two data sets).
These co-location criterion correspond to a trade-off between a small enough distance between TIRVIM and MCS data (in order to limit the effects of meteorological variability over these spatio-temporal scales), and a large enough  amount of co-located data.
Given the local time coverage of MCS data, this validation exercise will be based on "near-3~PM" and "near-3~AM" observations.
Several MCS profiles can match one TIRVIM profile, and conversely, a MCS profile can be found co-located with several TIRVIM profiles. 
As an example, for the near-3~PM observations in the first 45 sols period, this selection process results in 10,386 TIRVIM profiles being co-located with 6,201 unique MCS profiles.
Figure~\ref{fig:coloc_coverage} shows the coverage, as a function of latitude and local time, of all TIRVIM-MCS co-located observations for these first 45 sols. 
We note that TIRVIM data south of 55°S are mostly missing at all local times: those have been filtered out at the pre-processing stage and correspond to bad-quality flagged interferograms. Poor quality data at cold surface temperatures seem to be more frequent. This might reflect a poorer quality of the response of the instrument to low signal levels.
The longitude--latitude coverage of these two co-located data sets is illustrated in Fig.~\ref{fig:coloc_coverage} for the 3~PM data near Ls=150°, revealing a dense and nearly uniform coverage for both instruments.

In the following, we estimate the quality of the match between MCS and TIRVIM retrieved temperature profiles in several ways: 
\begin{itemize}
    \item We compute statistics based on MCS--TIRVIM pairs of co-located individual temperature profiles, for strict co-location criteria.
    \item We compare latitude-pressure sections of zonally-averaged temperatures, for strict co-location criteria.
    \item We also compare latitude-longitude sections of temperature at given pressure levels, acquired at similar dates and local times, but without considering spatial co-location.
\end{itemize}

\subsection{Comparison of individual temperature profiles}

\begin{figure}
  \centering
  \includegraphics[width=0.9\linewidth]{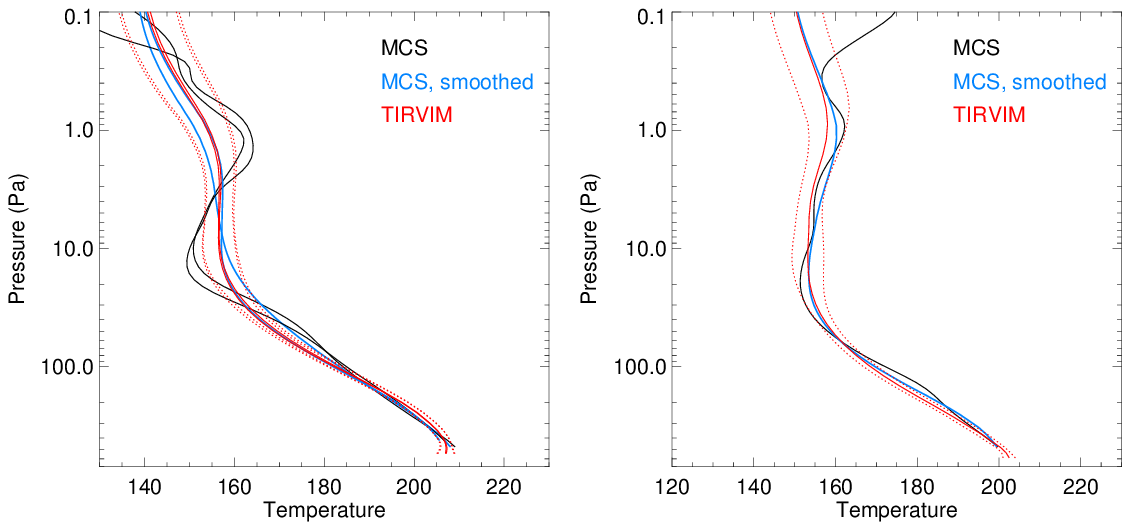}
  \caption{Retrieved temperature profiles from TIRVIM  (in red, with error envelop in dotted lines) compared with co-located MCS profiles (in black) and smoothed versions of the MCS profiles (in blue) taking into account TIRVIM averaging kernels and the influence of its \textit{a priori} profile, as described in the text.
  The example on the left is for latitude 45°N, the one on the right is for 65°N ; both were acquired near 3~AM. 
  Note that the vertical axis ranges till 0.1~Pa for context, to  display the behavior of MCS temperature profiles at higher altitudes, but TIRVIM data are sensitive only up to $\sim$2~Pa.}
  \label{fig:vertical_profiles_examples}
\end{figure}

\begin{figure}
  \centering
  \includegraphics[width=0.7\linewidth]{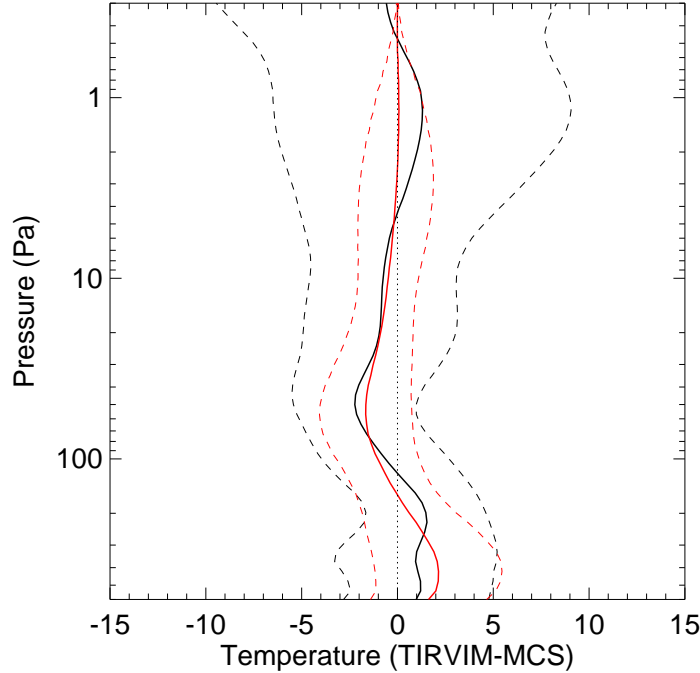}
  \caption{Statistics of the TIRVIM-$\mathbf{\hat{T}_\mathrm{MCS}}$ difference based on thousands of co-located temperature profiles near 3~AM, Ls$\sim$150°. The red solid line shows the average of this difference, the red dashed line is the 1-$\sigma$ standard deviation of this difference. Black lines are the same but considering the difference between TIRVIM and the raw MCS profiles.}
  \label{fig:all_vertical_profiles_3h}
\end{figure}

Comparison of TIRVIM and co-located MCS temperature profiles allows us not only to assess  potential biases between the two sets of retrievals, but also to evaluate the impact of the different vertical sensitivity in comparing a nadir with a limb sounder.
We present two selected cases of the direct comparison between TIRVIM and MCS temperature profiles in Figure~\ref{fig:vertical_profiles_examples}.
In one case, near 65°N and 3~AM, the comparison is excellent between these two instruments.
In the second example, a significant mismatch (up to 10K) is found over the range 1--10~Pa.
Here, the difference in vertical resolution and sensitivity is striking, with TIRVIM profiles being much smoother than those from MCS. 
In particular, TIRVIM does not capture the sharp vertical oscillation seen by MCS in the range 1--10~Pa. As mentioned in section~\ref{sec:algo}, this is consistent with the broad TIRVIM functional derivatives in that pressure range.
To further investigate this, we can apply TIRVIM averaging kernels to MCS profiles to emulate  the coarser vertical sensitivity of TIRVIM on MCS profiles  and to account for the fact that  when TIRVIM information content is low, the temperature smoothly goes back to an \textit{a priori} profile. 
In a fashion similar to the smoothing of the MCD profiles done in our OSSE described in section~\ref{subsec:synthe_temp}, we thus replace $\mathbf{T_\mathrm{true}}$ by  $\mathbf{T_\mathrm{MCS}}$ in equation~\ref{eq_lissage} to derive a smoothed MCS profile, noted $\mathbf{\hat{T}_\mathrm{MCS}}$.
This procedure follows the methodology for inter-comparing profiles retrieved from different instruments documented in \citeA{rodgers2003}.
By employing equation~\ref{eq_lissage} in such an inter-comparison exercise, we assume that the MCS profiles are not influenced by their own \textit{a priori} (which holds true on the pressure range considered here, 1--500~Pa) and that the MCS vertical resolution significantly exceeds that of TIRVIM, which is also satisfied. 
When we apply equation~\ref{eq_lissage}, the emulated $\mathbf{\hat{T}_\mathrm{MCS}}$ profiles are now much closer to the retrieved TIRVIM profiles (see Figure~\ref{fig:all_vertical_profiles_3h}), as can be expected. Individual differences that previously reached 10K in the direct comparison are now smaller than 3K.
In the following, we thus focus on the comparison between TIRVIM and the smoothed MCS profiles, as this is more relevant to validate our results.

We then compute statistics based on all pairs of MCS-TIRVIM co-located profiles.
Figure~\ref{fig:all_vertical_profiles_3h} displays the average and 1-$\sigma$ standard deviation  of the TIRVIM--$\mathbf{\hat{T}_\mathrm{MCS}}$ difference for all co-located profiles acquired near 3~AM. 
Both the bias and standard deviation of the difference tend to be zero at high altitudes, which simply results from the fact that the smooth MCS profiles go back to the same \textit{a priori} profile as the one used for TIRVIM retrievals. This tendency starts between 1 and 2~Pa, which is consistent with the information content in nadir observations and remind the fact that TIRVIM retrievals should not be interpreted at pressures lower than 1--2~Pa.
Over the range 2--500~Pa, the average of the difference between the two datasets is within $\pm$2K, which we take as an estimate of TIRVIM accuracy. The standard deviation of the TIRVIM--$\mathbf{\hat{T}_\mathrm{MCS}}$ difference is in the range to 2--3~K, which is then an estimate of TIRVIM precision.
These figures are of the order of magnitude of the typical formal retrieval error of either TIRVIM or MCS temperatures. 
Hence, this shows that when differences in vertical sensitivities are taken into account, TIRVIM and MCS are in excellent agreement.
For completeness, we include in Figure~\ref{fig:all_vertical_profiles_3h} the same statistics should the raw MCS temperature profiles be taken in consideration. This illustrates the precision that would be (wrongly) derived if one were to disregard differences in vertical resolution between instruments.
Similar results are obtained for the 3~PM case.

  \subsubsection{Comparison of the zonally-averaged temperature}

We now focus on the comparison of the latitude-pressure thermal structure from TIRVIM and MCS near 3~AM and 3~PM, considering zonal averages of the co-located measurements, in bins of 5° latitude.
One word of caution regarding the computation of zonal averages: while TIRVIM (nadir) retrievals are sensitive to almost down to the surface (except in extreme dusty or cloudy conditions not encountered here), MCS profiles often display a limited vertical coverage, extending only down to 80~Pa (20~km), for instance. This occurs when the optical depth of water ice and/or dust was too high at lower tangent altitudes to allow a reliable temperature retrieval from the MCS limb data. 
We thus have to be careful and avoid comparing a TIRVIM zonal average obtained from a rather uniform longitudinal sampling with a MCS "zonal average" for which the longitudinal coverage could be uneven and vary with altitude. 
For each 5°-wide latitudinal bin considered, we thus proceed by browsing one by one the MCS pressure grid and search for co-located MCS-TIRVIM data only among a subset of the MCS database, where temperature was indeed retrieved at that level, before averaging the temperatures. In the following, for simplicity, we call zonal average the average over observed longitudes.

\begin{figure}
  \centering
  \includegraphics[width=0.95\linewidth]{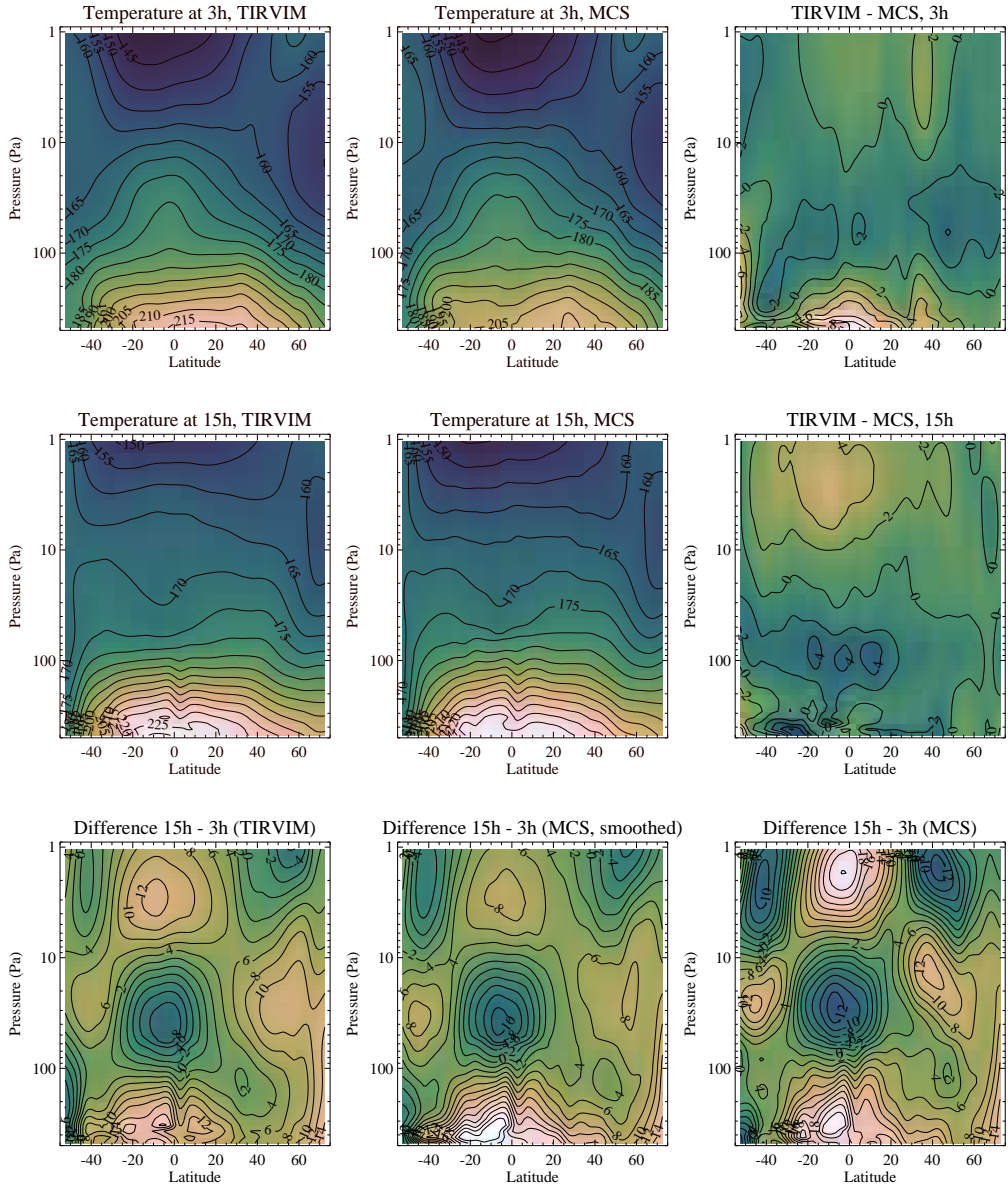}

  \caption{Pressure-latitude sections of the zonally-averaged temperature near 3~AM (top), 3~PM (middle), and the difference between 3~PM and 3~AM (bottom). In the two top rows, left panels correspond to TIRVIM, middle panels to $\mathbf{\hat{T}_\mathrm{MCS}}$ and the difference between the two datasets is shown in the right panels. In the bottom row, TIRVIM results are shown on the left, smoothed $\mathbf{\hat{T}_\mathrm{MCS}}$ in the middle and raw MCS ones on the right. These figures highlight the diurnal mode of martian thermal tides. Bins of 5° wide in latitude are used. In each panel, zonal averages are performed for a subset of the available TIRVIM and MCS data that meet co-located criteria, as described in the text.}
  \label{fig:MCS_ACS}
\end{figure}

The resulting latitude-pressure cross-sections of the zonal-mean temperature are shown in Figure~\ref{fig:MCS_ACS} for TIRVIM and $\mathbf{\hat{T}_\mathrm{MCS}}$ for 3AM and 3PM, along with the cross-sections of the TIRVIM--$\mathbf{\hat{T}_\mathrm{MCS}}$ difference. 
Similar figures for MCS temperature (on their own retrieval pressure grid, without vertical smoothing) are shown in Figure~\ref{fig:MCS_ACS_supp} in Supplementary Material.
We also display the day/night temperature difference (3~PM$-$3~AM) for TIRVIM, $\mathbf{\hat{T}_\mathrm{MCS}}$ and MCS in the bottom panels of Figure~\ref{fig:MCS_ACS}.
The agreement between the two data sets is very satisfactory, both qualitatively and quantitatively. Before applying the averaging kernels to MCS profiles, temperature differences between TIRVIM and MCS data sets are already mostly in the range $\pm 4$~K except at high altitudes (pressures lower than 10~Pa) where it can reach 12K. These larger discrepancies between TIRVIM and MCS can be explained by differences in their inherent vertical sensitivity, as discussed previously.
Indeed, when comparing TIRVIM and $\mathbf{\hat{T}_\mathrm{MCS}}$, these differences decrease and become smaller than 4K everywhere, except at 3AM in the lower atmosphere. 
The TIRVIM--$\mathbf{\hat{T}_\mathrm{MCS}}$ temperature differences at pressures greater than 200~Pa at 3~AM is not well understood and may partly reflect an artefact of applying equation~\ref{eq_lissage} on the lower portion of MCS profiles and/or partly result from an actual bias of either TIRVIM or MCS retrievals at these pressures.

The main features of the temperature field seen by MCS are well captured by TIRVIM, both at 3~AM and 3~PM. 
At this season (end of summer in the northern hemisphere) and in the  lower atmosphere (pressures greater than 200~Pa), the meridional temperature structure is characterized by warmer temperatures in the northern hemisphere compared to the southern hemisphere. 
The temperature decreases rapidly towards high southern (winter) latitudes, from typically 210K at 30°S to 180K at 50°S, at the edge of the cold polar vortex.
At lower pressures, the meridional thermal structure is more symmetric about the equator, as the dominant feature is the thermal tide pattern.

The diurnal mode of the thermal tide is well visible in the day/night temperature difference shown in the bottom panels of Figure~\ref{fig:MCS_ACS}.
These figures highlight the well-known pattern of positive and negative temperature anomalies vertically stacked, at the equator and 45° latitude (with an opposite sign). 
These patterns have been well documented in the past, for instance by \citeA{Lee2009} and \citeA{Kleinbohl2013} from the analysis of MCS data. 
TIRVIM captures well the signs and locations of these temperature extrema, and the amplitude of these extrema are similar to that of the smoothed MCS temperature profiles ($\mathbf{\hat{T}_\mathrm{MCS}}$). For instance, near the equator, the 30~Pa afternoon temperature is found by TIRVIM to be $\sim$8~K colder than at night (10K in the $\mathbf{\hat{T}_\mathrm{MCS}}$ field), and the 3~Pa afternoon temperature is found to be $\sim$12~K warmer than at night (8K in the $\mathbf{\hat{T}_\mathrm{MCS}}$ field). 
The actual amplitudes of the thermal tides are known to increase even more with decreasing density, as can be seen in the raw MCS day-night temperature difference in Figure~\ref{fig:MCS_ACS}.
Indeed,  in the raw MCS data, the amplitude of the 3~PM$-$3~AM difference reaches +20K at the equator at 2~Pa, and +10--12K near 45° latitude at 2~Pa as well. These large extrema value are linked with sharp vertical gradients or oscillations of the temperature profile at these altitudes (see for instance the local temperature minimum seen in the MCS profiles at 3~AM and 45°N in Figure~\ref{fig:MCS_ACS_supp}) that TIRVIM cannot capture properly.
The fact that TIRVIM provides a muted version of the actual thermal tide signal at 1--10~Pa despite a moderate sensitivity in this region and a growing influence of the \textit{a priori} profile (which is not based on any prior knowledge on how the thermal tide should look like) complies with its coarse vertical resolution -- as shown from the good comparison with $\mathbf{\hat{T}_\mathrm{MCS}}$ -- and is thus very satisfactory.

\subsubsection{Comparison of longitude--latitude sections \label{sec:compa_latlon}}

\begin{figure}
  \centering
  \includegraphics[width=0.9\linewidth]{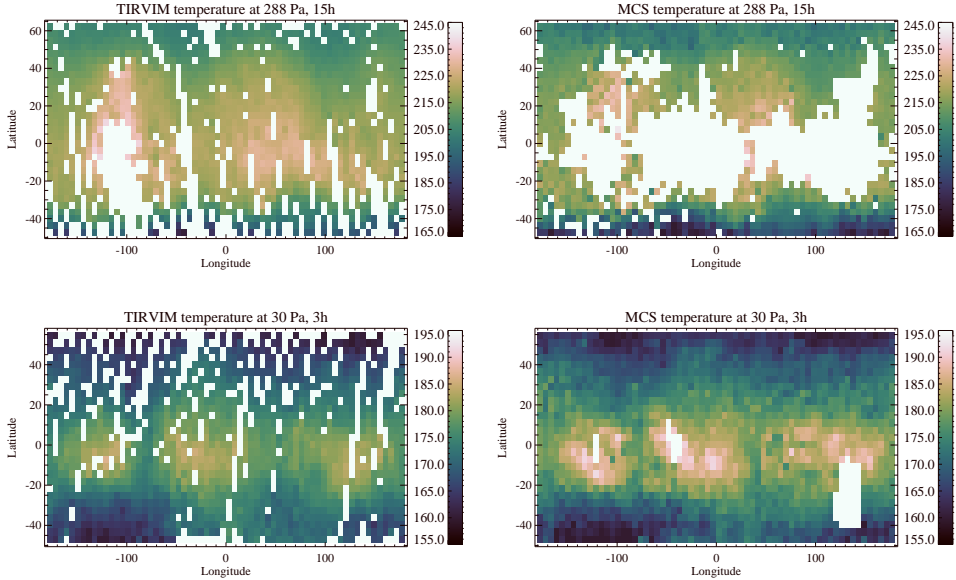}
  \caption{Temperature maps as derived from TIRVIM (left column) and provided by MCS (right column) at Ls$\sim$150°, at 3~PM and 288~Pa (upper row) and at 3~AM and 30~Pa (lower row). Both TIRVIM and MCS feature similar longitudinal wave patterns.}
  \label{fig:MCS_ACS_lat_lon}
\end{figure}

We also investigated the robustness of the retrieved temperature field in terms of longitudinal variations.
Here we consider the retrieved temperature binned in latitude and longitude with bin width of 3.75°$\times$5.625°, a resolution typical of GCM simulations or the MCD.
We consider all TIRVIM or MCS data acquired near 3~AM or 3~PM ($\pm$1~h) and restrict the date between Ls=149° and 155° (sols 316--329). 
Hence, we do not impose strict co-location between the two datasets, except for similar season and local times, and each of the resulting temperature map reflects the inherent latitude/longitude coverage of each instrument.
We detail below two examples: the longitude/latitude cross-section of temperature at 288~Pa, 3~PM and that at 30~Pa, 3~AM, shown in Figure~\ref{fig:MCS_ACS_lat_lon}.

Firstly, we highlight that MCS coverage features significant gaps in the equatorial region regarding the afternoon temperature at 288~Pa. This is linked to enhanced cloud opacity, as this corresponds to the end of the aphelion belt season, which hampers limb retrievals. TIRVIM has the advantage to provide a more uniform coverage at these altitudes.
A prominent wavenumber-3 feature is seen at 30~Pa and 3~AM, well captured by TIRVIM. 
The better coverage of TIRVIM data in the lower atmosphere allows a more precise characterization of a similar wave seen at 3~PM, 288~Pa. 
As these wave signatures are visible in a constant local time framework, this indicates that they are non-migrating thermal tides. An apparent wavenumber 3 can correspond to the diurnal Kelvin wave with zonal wavenumber 2, already reported eg. by \citeA{Banfield2003, Wilson2000} from TES and \citeA{Guzewich2012, Forbes2020} from MCS observations.
The analysis of these waves and of the 4-D structure of the thermal tides is deferred to future work. The local time coverage of TIRVIM will be an asset to better characterize, for instance, the amplitudes and phases of different modes of the thermal tides.

Hence, we conclude that TIRVIM temperature retrievals are reliable, whether zonally-averaged temperature fields or individual temperature profiles are considered, as long as their coarse vertical resolution are kept in mind. 
The main features of the diurnal mode of the thermal tides and longitudinal waves are well reproduced, compared to MCS observations.
This validation exercise performed near 3~AM and 3~PM is a necessary step before the scientific exploitation of temperature variations at all local times, which will be addressed in subsequent publications.

\subsection{A first assessment of dust total column opacity retrievals \label{sec:results:dust}}

\begin{figure}
  \centering
  \includegraphics[width=0.7\linewidth]{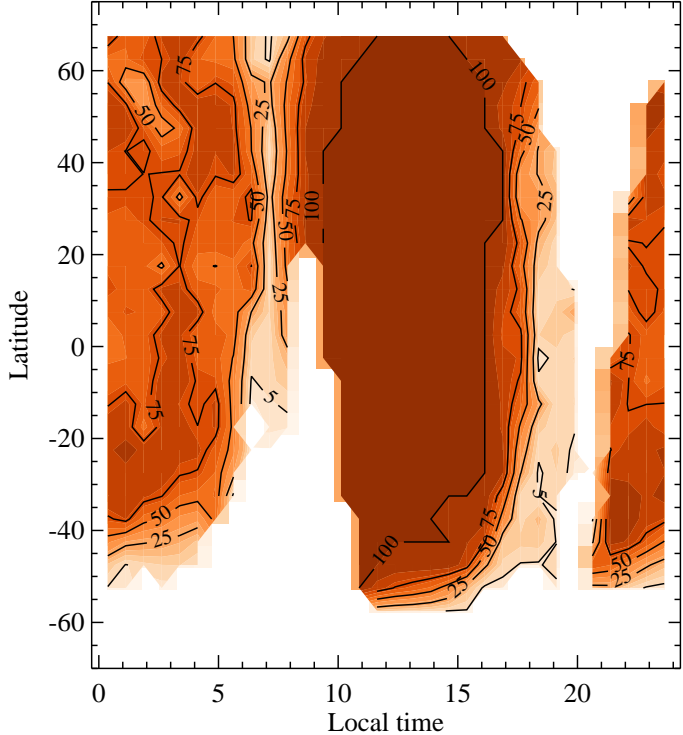}
  \caption{Percentage of retrievals that pass our quality flag for dust retrievals, for all first 45 sols of TIRVIM observations (Ls 142° to 167°), as a function of latitude and local time. White areas are missing data (the latitude - local time coverage after 45 sols is not complete).}
  \label{fig:fraction_qflag_dust}
\end{figure}

In section~\ref{sec:inv:aero}, we have defined a quality filter for dust that is satisfied if the derivative of the radiance at 1100~cm$^{-1}$ over a relative change in dust opacity is greater than the 1-$\sigma$ NER estimated at 1100~cm$^{-1}$. In other words, we retain dust opacity values for which the estimated error is less than 100\%.
The fraction of dust retrievals that passed this quality (or sensitivity) filter is shown as a function of local time and latitude in figure~\ref{fig:fraction_qflag_dust}, for all TIRVIM data acquired between Ls=142° and 167°.
As expected and already noted in section~\ref{sec:synthe}, two blind zones exist near 7AM and 7PM when retrieving dust is not possible due to the lack of sufficient temperature contrast between the surface and the lower atmosphere, where dust lies.
During daytime (10AM -- 5PM), 100\% of dust retrievals pass our quality filter due to a high sensitivity to dust (a combination of warm surface and cold atmosphere) except at high southern winter latitudes, where there is a low surface/atmosphere thermal contrast even during daytime.
At nighttime, most dust retrievals (60--80\%) pass our quality filter, however, the uncertainty associated with those nighttime retrievals is quite large, as we will see later.

\begin{figure}
  \centering
  \includegraphics[width=0.7\linewidth]{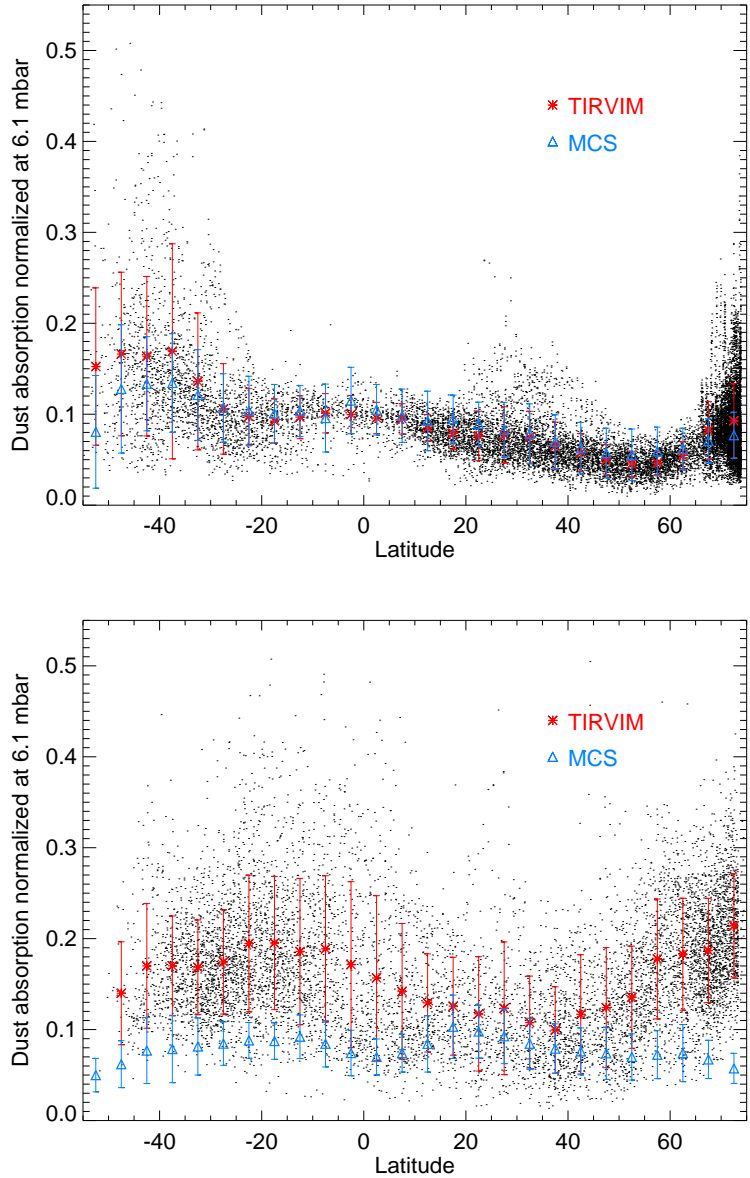}
  \caption{Retrieved dust absorption opacity from TIRVIM near 3~PM (top) and 3~AM (bottom) as a function of latitude, for Ls$\sim$150°. Black dots stand for individual retrieved values, while red stars indicate zonal averages for 5°-wide latitude bins. For validation purpose, blue triangles are zonal averages of MCS retrieved dust absorption scaled at 9.3~$\mu$m, for data co-located with TIRVIM measurements. Vertical bars illustrate the variance of each  data set within a latitudinal bin.}
  \label{fig:dust_validation}
\end{figure}

Validation of the retrieved dust opacity is evaluated near 3~AM and 3~PM by comparison with co-located MCS data, with the same co-location criteria as for the validation of atmospheric temperature. 
We exploit the data product called \textit{coldust} provided by the MCS team, which is total dust extinction at 21~$\mu$m obtained from extrapolation of the dust profile down to the surface. This extrapolation was done under the assumption that dust is well-mixed below the level of the last valid measurement \cite{Montabone2020}.
We multiply this value by 2.7 to estimate the MCS dust extinction at 9.3$\mu$m and further divide it by 1.3 to get from dust extinction to dust absorption, as recommended by \citeA{Montabone2015, Montabone2020} and based on the work by \citeA{Smith2004, Wolff2003}. Indeed, we recall that we neglect scattering effects and only derive an effective dust absorption.
Dust absorption derived from TIRVIM and MCS data are compared in figure~\ref{fig:dust_validation} for daytime and nighttime observations during the first 45 sols. 
Overall, TIRVIM daytime retrievals are in excellent agreement with MCS, and the associated 1-$\sigma$ standard deviation of that difference for individual dust opacity retrievals (not shown) is about 10\%. Dust opacity (in absorption, at 9.3$\mu$m) is low, of the order of 0.1, which is expected at this season.

At night, we note a significant scatter in the derived dust opacities from TIRVIM. This is actually consistent with the rather poor S/N ratio and individual errors that are near 40\% (hence a 1-$\sigma$ error of 0.08 for a 0.2 dust opacity). 
More worrying is a nearly systematic overestimation of nighttime dust opacity by TIRVIM, except at latitudes 15°N--40°N.
The reason behind this bias is highly similar to one of the cases described in our OSSE and illustrated in figure~\ref{fig:synthe6}: our  assumed profile for dust vertical distribution overestimates the dust loading above the first scale height, which happens to be at the same atmospheric temperature as the surface.
Indeed, we compared the MCS vertical profiles of dust opacity to our assumed one and confirms that the observed dust (by MCS) decreases more steeply with altitude above the first scale height. 
In this example, we are blind to dust located near 1.5 scale height (as it emits at a similar brightness temperature as the surface) and we are mostly sensitive to dust in the first scale height above the surface (where the surface-atmosphere temperature contrast is high enough).
As we only derive a scaling factor to a profile that has too much dust at higher altitudes, this results in an overestimated total opacity.

To overcome this problem, we will have to improve our assumptions about the vertical distribution of dust in a future version of our algorithm.
This is not trivial, as previous studies have shown that the dust vertical distribution can vary quite dramatically with latitude, season, and with local time in MCS data \cite{Heavens2011,Kleinbohl2020}.
As MCS only provides constraints near 3~AM and 3~PM, it appears difficult to set up a robust vertical profile that would be relevant for any local time observed by TIRVIM.
However, as we have previously shown that our retrieval performs well during the day (even if our assumed vertical distribution is wrong), and because dust retrievals from a nadir sounder are not possible at evening and morning times anyway, it might be sufficient to assume the nighttime MCS profile as a baseline for all our retrievals. This will be investigated in future work.


\section{Summary and Conclusions\label{sec:discuss}}

In this paper, we have documented and evaluated a retrieval algorithm tailored to ACS/TIRVIM thermal infrared spectra acquired in nadir viewing geometry, with the goal of providing the best estimates of Mars' surface temperatures, vertical profiles of its atmospheric temperature (between 5 and 50~km) and the integrated infrared optical depths of dust and water ice clouds.

This algorithm was first tested on a set of synthetic spectra in conditions chosen to be representative of the variability of the martian atmosphere at various seasons, latitudes and local times extracted from the Mars Climate Database. This constitutes our Observing System Simulation Experiment (OSSE).
The precision and accuracy of these synthetic retrievals were carefully evaluated, and in parallel, the information content of the data was studied in detail.
Regarding surface temperature and aerosol opacity retrievals, we find that our algorithm performs very well for the most favorable scenes, i.e. scenes with a warm surface temperature and featuring a large contrast between the surface and the atmospheric layer where aerosols reside.
We caution that there are frequent conditions when this temperature contrast is low, for which the optical depth of aerosols cannot be constrained. This mostly (but not solely) occurs for observations acquired near dawn and dusk.
For intermediate cases, for which we have a moderate sensitivity to dust and/or clouds, the retrievals can be significantly biased due to wrong assumptions on the vertical distribution of aerosols and/or a wrong estimate of the temperature in the lower atmosphere. 
Uncertainty values on dust optical depth are highly variable. They are of the order of 5-20\% when the surface-atmosphere temperature contrast is high (typically, daytime observations) and of 20 to 60\% when this contrast is moderate (nighttime conditions).
Error on retrieved surface temperature is of the order of 1K for warm surfaces ($>$220K), and 3K for colder surfaces. However, significant biases (up to 10K) are reported, in particular linked with a wrong determination of retrieved dust or ice opacity, as there are situations where there is a high level of degeneracy between these quantities.
Future work should thus focus on a more realistic representation of the vertical distribution of dust and water ice clouds in our algorithm, either with the help of models or other independent observations.
In order to improve our aerosol estimates, we also envision in the future to co-add spectra acquired at close locations (in particular for low surface temperatures) to increase their signal-to-noise ratio.

The retrieval of atmospheric temperature is found to be very robust. Most of the differences found between retrieved temperatures and the input temperatures used to generate synthetic spectra in our OSSE can be explained by the coarse vertical resolution of TIRVIM, especially in the 2--10~Pa range where the width of the functional derivatives is of the order of two scale heights. These differences can locally reach 10K.
However, when TIRVIM vertical resolution is taken into account (through the use of averaging kernel matrices), our retrieved temperatures agrees with the input ones within 2--3K, which is taken as an estimate of uncertainty in our retrieved profiles and supports the good performance of our algorithm.  
We report some difficulties in retrieving atmospheric temperatures in the first scale height when there are optically thick clouds at low altitudes. These thick clouds limit our sensitivity to the lower atmosphere as the result of the degeneracy between surface temperature, cloud opacity and/or low atmospheric temperature retrievals. 
Important biases are also found when there is a very steep temperature inversion in the first scale height, also linked with the rather coarse vertical resolution of TIRVIM. However, these cases remain rare.

We then applied our algorithm to the first 45 days of TIRVIM operations, corresponding to Ls=142--167° of Martian Year 34.
Retrieved temperature profiles were validated against co-located measurements by the Mars Climate Sounder near 3~AM and 3~PM. 
As for the OSSE, we find that most of the differences between the thermal structure derived from TIRVIM and MCS can be attributed to differences in vertical resolution and sensitivity. 
Local temperature extrema (vertical oscillations of the temperature profile) seen by MCS at pressures lower than 10--20~Pa are not well captured by TIRVIM, consistently with its coarse vertical resolution. 
As a consequence, the signature of the diurnal mode of the thermal tide, well visible in the day-night temperature difference, appears severely muted in TIRVIM retrievals at these low pressure levels.
On the positive side, TIRVIM retrievals are highly comparable to MCS temperatures in the range 300--20~Pa (within 2--3K). Furthermore, they have the advantage of a more homogeneous spatial coverage at higher pressures, where MCS is frequently blind due to larger aerosol opacity along the limb-viewing light path.
We also showed that several wave patterns are well visible in longitude-latitude cross-section of the retrieved temperature from TIRVIM. 

These results are promising and open the way to future detailed studies of atmospheric dynamics at the diurnal scale in Mars' lower atmosphere.
Indeed, the strength of the TIRVIM dataset lies in its sampling of the diurnal cycle, with all local times being sampled on the planet every 54 sols. As an example, full local time coverage is an asset to better characterize the amplitudes and phases of the different modes of the thermal tides.


\appendix

\section{Supplementary figures}

\subsection{Quality of dust retrievals}

\begin{figure}[!b]
  \centering
  \includegraphics[width=0.9\linewidth]{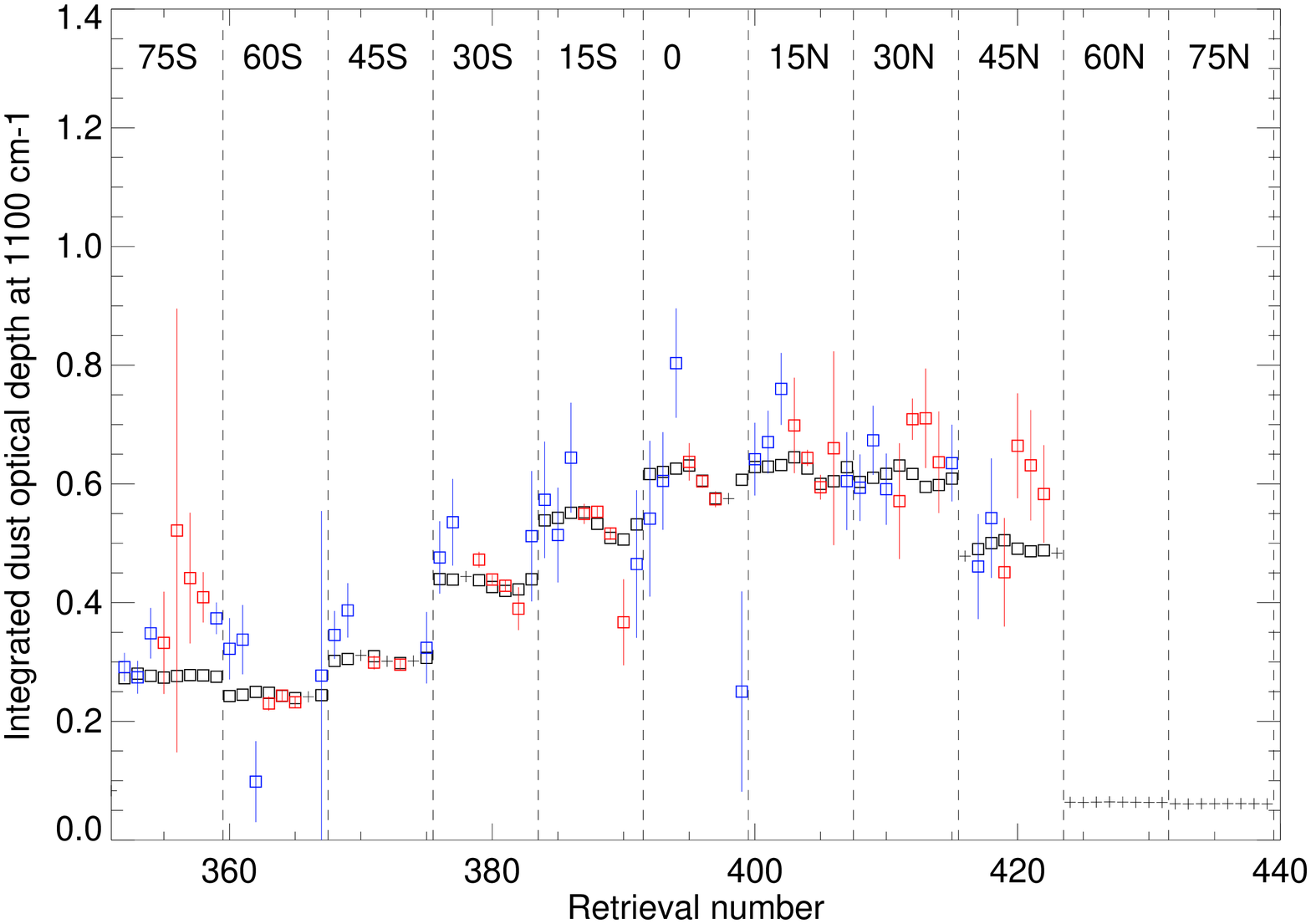}
  \caption{Same as Figure~\ref{fig:dust_mcd_all} but only for data at Ls=270°, with labels for different latitudes. For a given latitude, data is sorted by local time (from midnight to 9~PM, every three hours).
  No satisfactory dust retrievals were obtained at latitudes 60N and 75N (winter) due to cold surface and poor signal-to-noise ratio.}
  \label{fig:dust_mcd_270}
\end{figure}

Figure \ref{fig:dust_mcd_270} shows the "true" versus retrieved dust opacity at different latitudes and local times, highlighting a subset of Figure~\ref{fig:dust_mcd_all} in the main text.

\subsection{Examples of synthetic retrievals}

In this appendix we display five detailed examples (out of the 440 test case retrievals of our OSSE) showing fits to synthetic TIRVIM spectra ; functional derivatives (Jacobian) for the retrieval of atmospheric temperature ; and the comparison between "true" and retrieved quantities. 
These examples are chosen as they exhibit significant errors on dust opacity, water ice opacity, surface temperature and/or atmospheric temperature despite satisfactory fits to the spectra. 
These figures are discussed in the main text.

\begin{figure}
  \centering
  \includegraphics[width=0.9\linewidth]{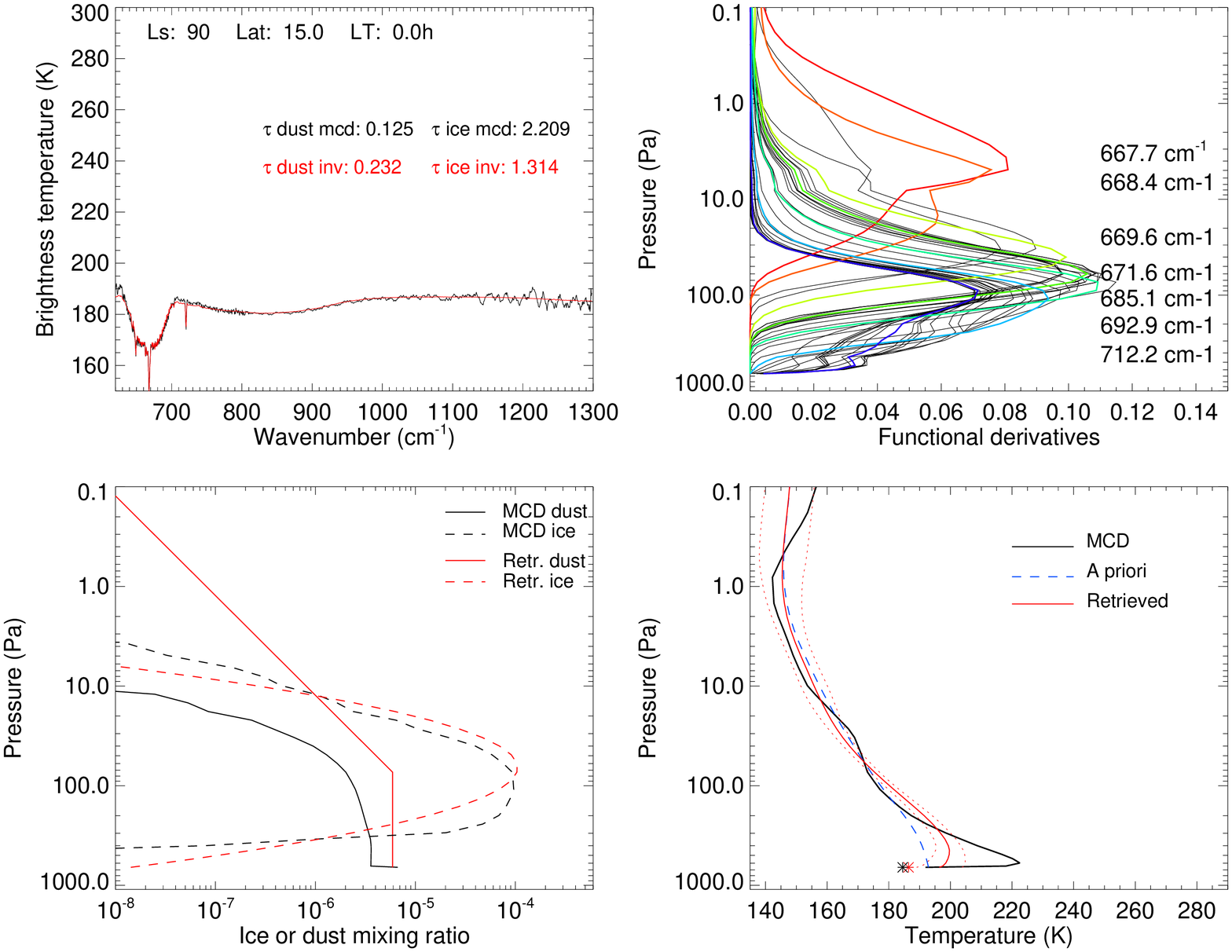}
  \caption{Example of a synthetic retrieval for a MCD scenario extracted at L$_s$=90°, latitude 15°N, local time 0~AM. Top, left: Synthetic TIRVIM spectrum (black) along with the best fit (red). Retrieved and MCD dust and water ice cloud integrated opacities are indicated. Top, right: functional derivatives of the temperature as a function of pressure, for 50 different wavenumbers within the CO$_2$ band. Several wavenumbers are highlighted in different colors and labeled. Bottom, left: Mixing ratio vertical profiles for dust and water ice as taken from the MCD (used to generate the synthetic spectrum, in black) and as derived from the retrieval process (in red ; note that only a scaling factor to a generic \textit{a priori} profile is retrieved). Bottom, right: Temperature vertical profile from the MCD (used to generate the synthetic spectrum, in black), \textit{a priori} profile built from the synthetic spectrum (dashed blue line), and retrieved profile (red), with error bars in red dotted lines. The black and red stars stand for the MCD and retrieved surface temperature, respectively.}
  \label{fig:synthe5}
\end{figure}

\begin{figure}
  \centering
  \includegraphics[width=0.9\linewidth]{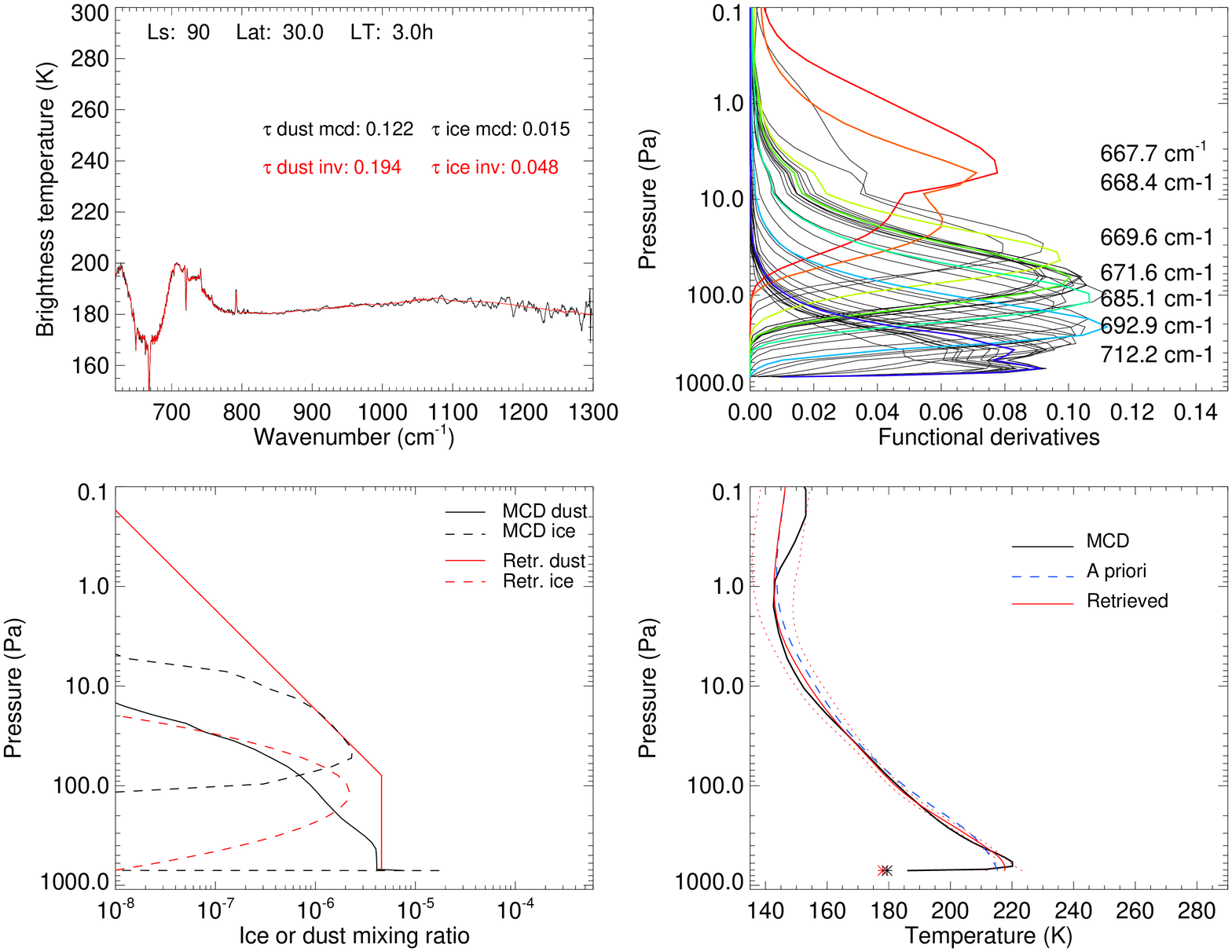}
  \caption{Same as figure~\ref{fig:synthe5} for  L$_s$=90°, latitude 30°N, local time 3~AM.}
  \label{fig:synthe6}
\end{figure}

\begin{figure}
  \centering
  \includegraphics[width=0.9\linewidth]{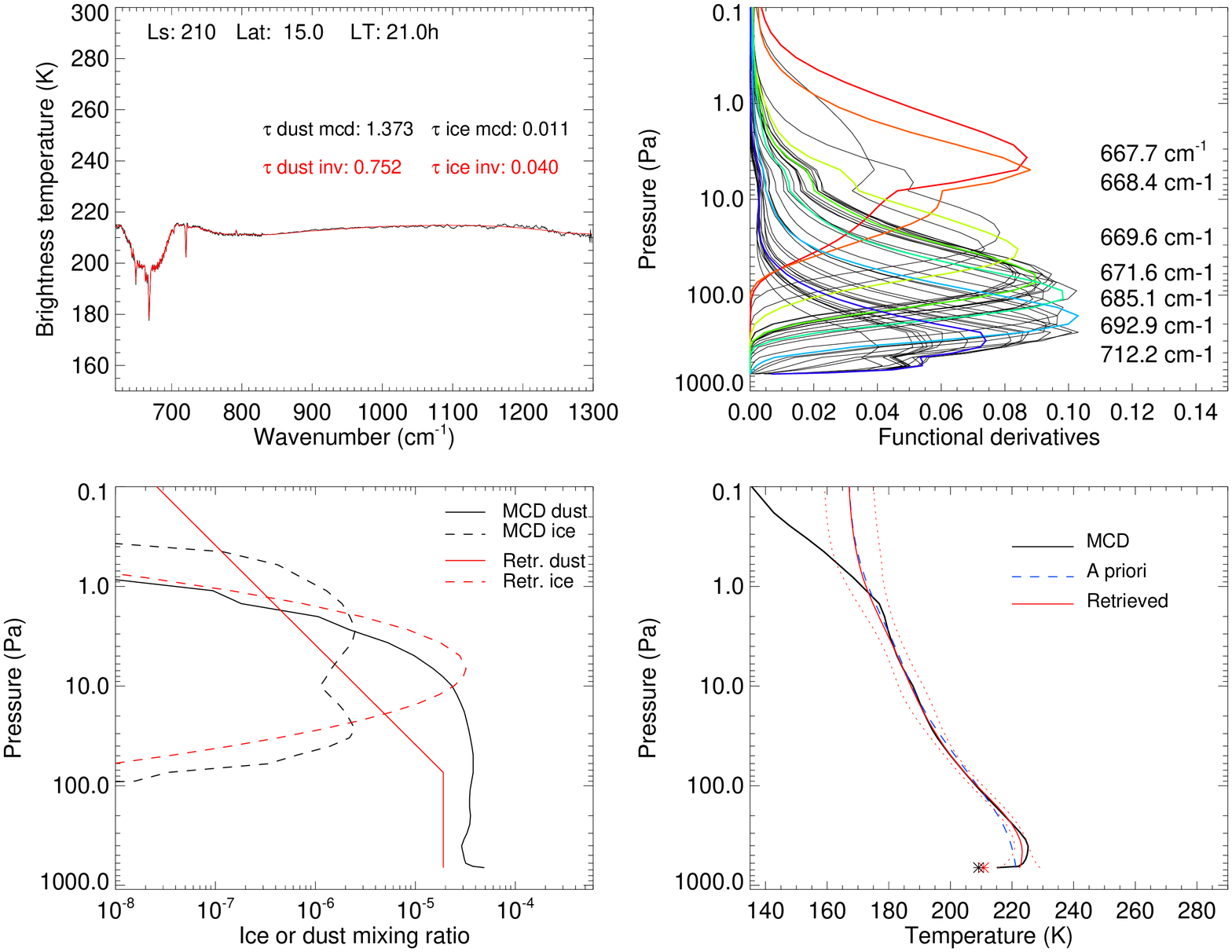}
 \caption{Same as figure~\ref{fig:synthe5} for  L$_s$=210°, latitude 15°N, local time 9PM.}
  \label{fig:synthe3}
\end{figure}

\begin{figure}
  \centering
  \includegraphics[width=0.9\linewidth]{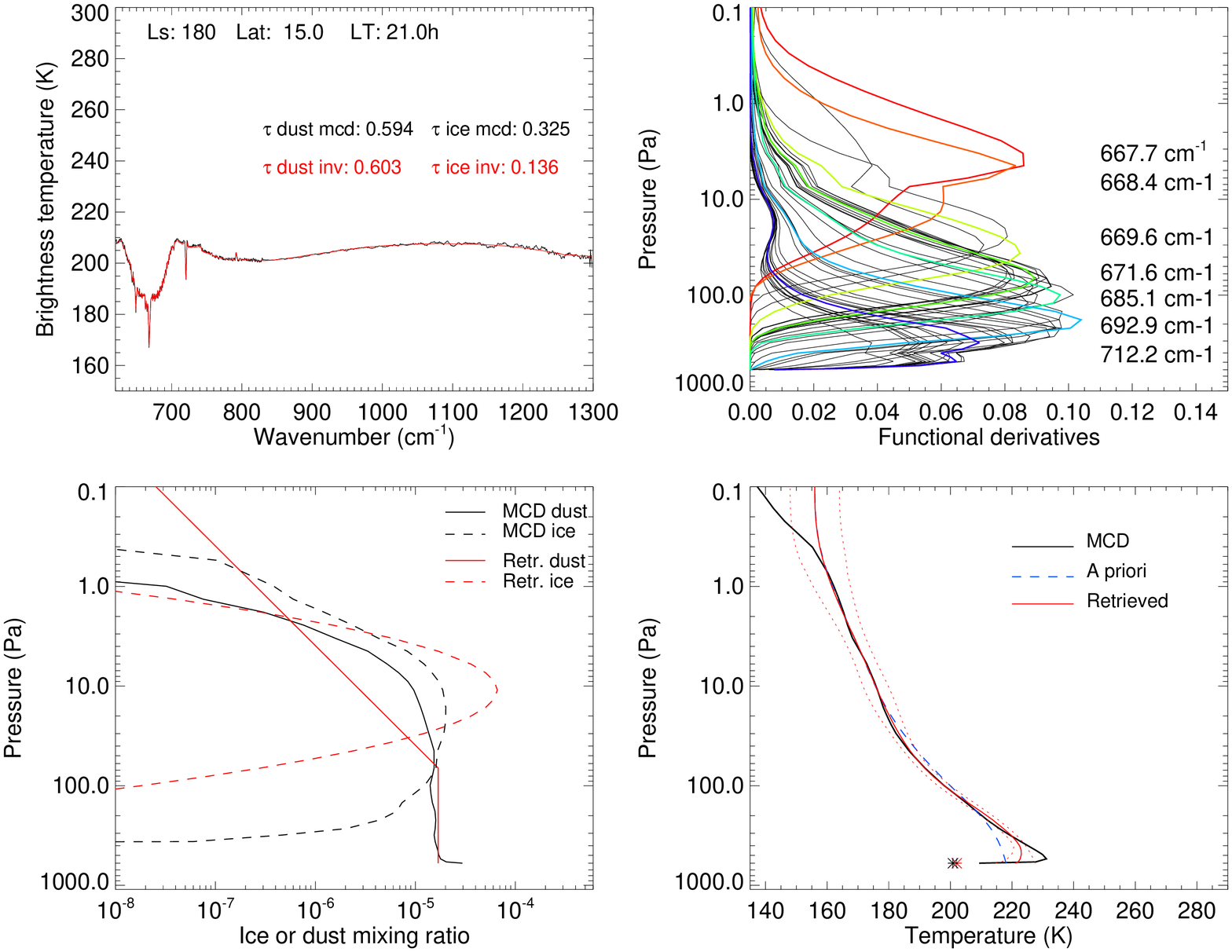}
  \caption{Same as figure~\ref{fig:synthe5} for  L$_s$=180°, latitude 15°N, local time 9~PM.}
  \label{fig:synthe2}
\end{figure}

\begin{figure}
  \centering
  \includegraphics[width=0.9\linewidth]{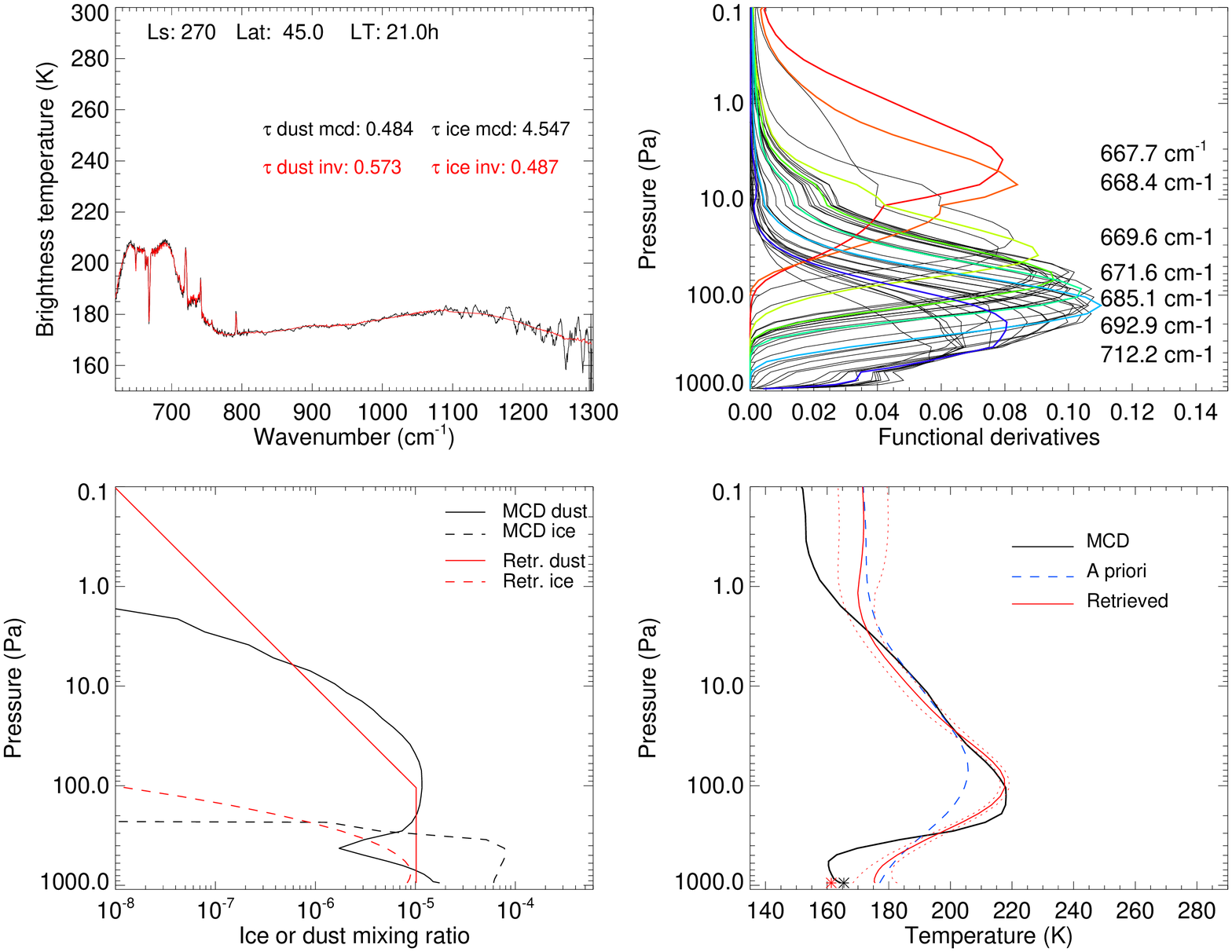}
  \caption{Same as figure~\ref{fig:synthe5} for a MCD scenario extracted at L$_s$=270°, latitude 45°N, local time 9PM.}
  \label{fig:synthe4}
\end{figure}


\subsection{Impact of using different \textit{a priori} temperature profile}

\begin{figure}
  \centering
  \includegraphics[width=0.9\linewidth]{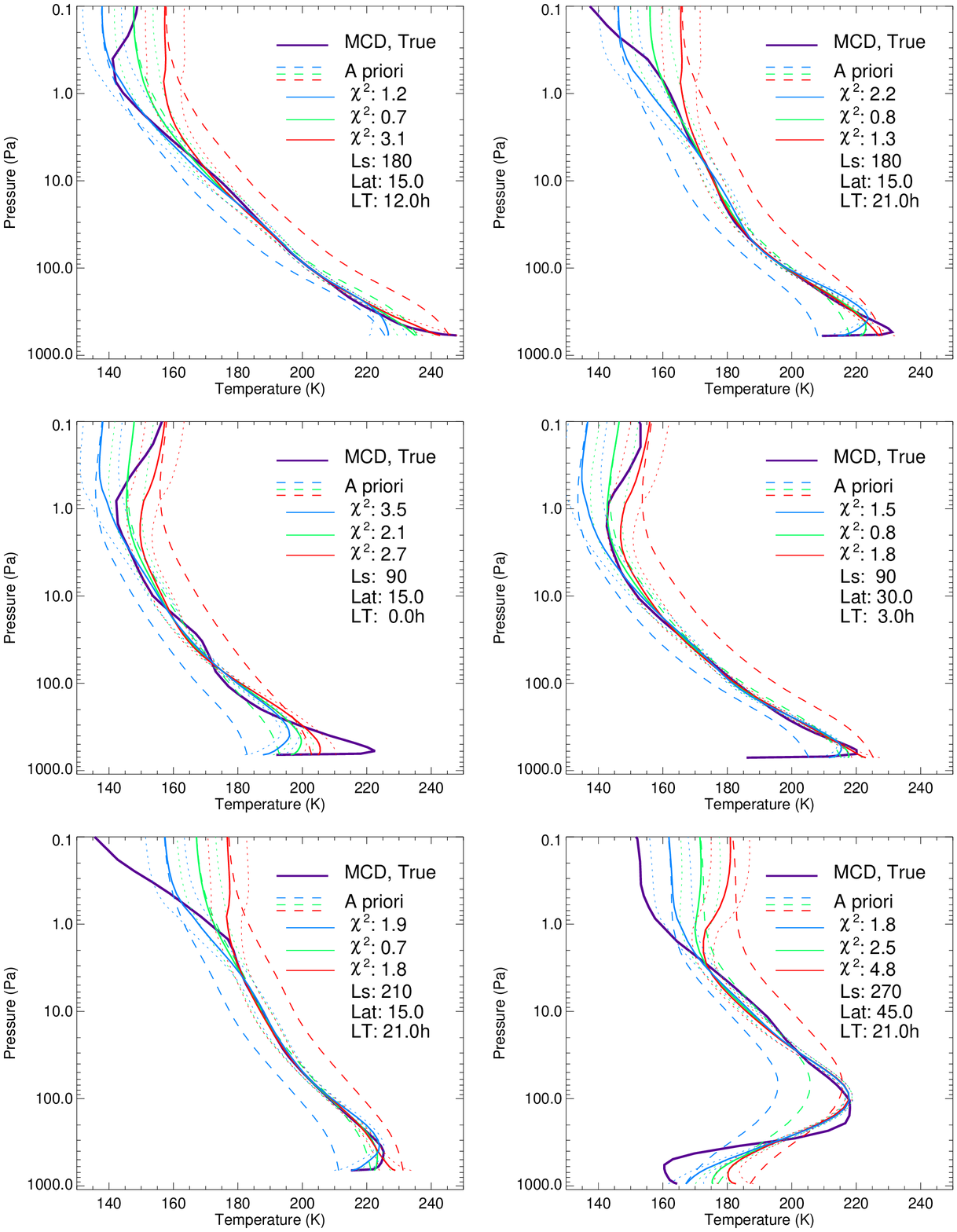}
  \caption{Retrieved temperature profiles (color solid lines) for different \textit{a priori} profiles (dashed lines): in green our nominal \textit{a priori} profile, in blue a profile 10K colder, in red a profile 10K warmer. Quality of the fit in the CO$_2$ band is given for each retrieval ($\chi^2$). These examples are for 6 out of 440 synthetic retrievals of our OSSE, with the "true" temperature profile in purple. Dotted lines highlight the error envelops. These six examples correspond to local times, season and latitudes of the cases shown in Figures~\ref{fig:synthe1} and \ref{fig:synthe5}--\ref{fig:synthe4}.}
  \label{fig:supp_prior}
\end{figure}

Figure~\ref{fig:supp_prior} demonstrates the good performance of our atmospheric temperature retrieval starting from different  \textit{a priori} profiles (see main text).

\subsection{TIRVIM-MCS comparison}

\begin{figure}
  \centering
  \includegraphics[width=0.95\linewidth]{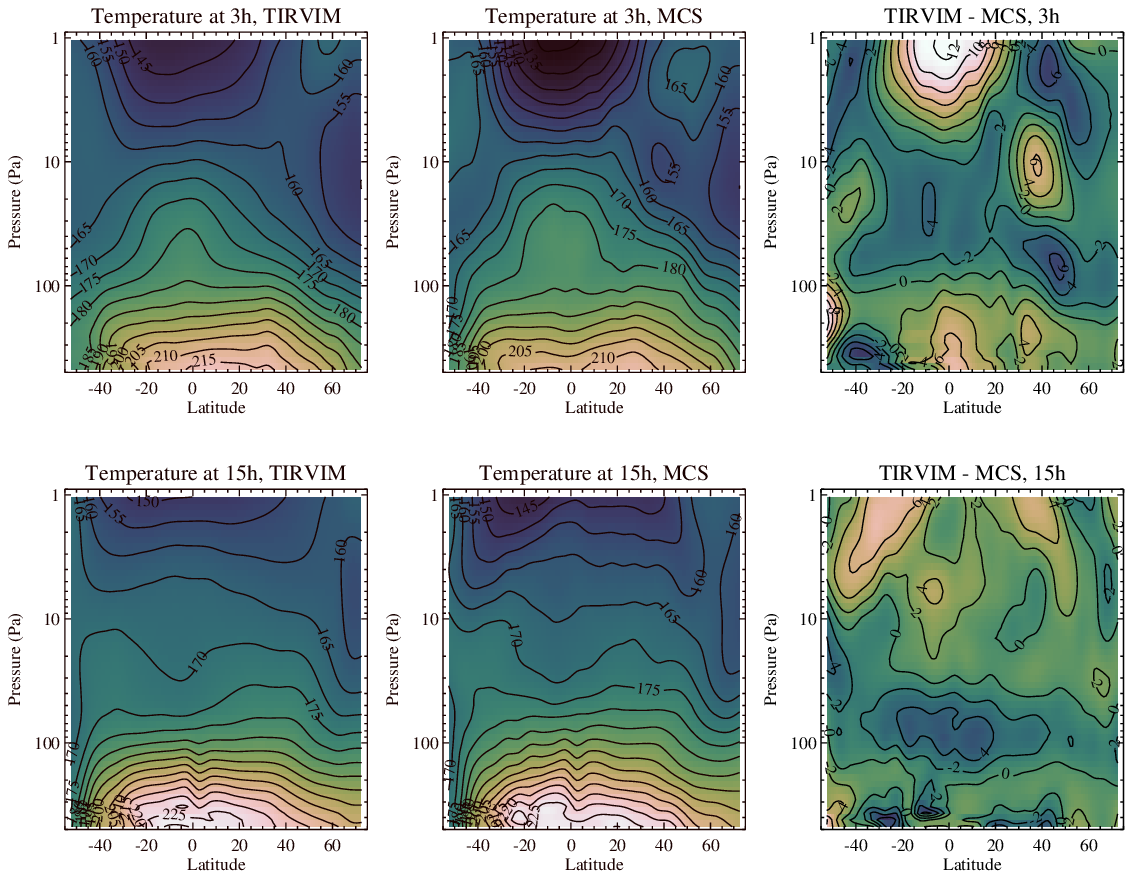}

  \caption{Pressure-latitude sections of the zonally-averaged temperature near 3~AM (top) and 3~PM (bottom). Left panels correspond to TIRVIM, middle panels to MCS and the difference between the two datasets is shown in the right panels. Bins of 5° wide in latitude are used. In each panel, these zonal averages are performed for a subset of the available TIRVIM and MCS data that meet co-located criteria, as described in the text.}
  \label{fig:MCS_ACS_supp}
  \end{figure}

Figure~\ref{fig:MCS_ACS_supp} shows the comparison of the zonally-averaged temperature near 3~AM and 3~PM as retrieved from TIRVIM and MCS, for co-located measurements.
No averaging kernels were applied to the MCS temperature fields in that figure.

%
%
%
%

%
%
%
%
%
%
%
%

\newpage
\acknowledgments
ExoMars is a space mission of ESA and Roscosmos. The ACS experiment is led by IKI, the Space Research Institute in Moscow, assisted by LATMOS in France. This work, exploiting ACS/TIRVIM data, acknowledges funding by CNES. The science operations of ACS are funded by Roscosmos and ESA. RMBY acknowledges support from UAE University grants G00003322 and G00003590. ACS/TIRVIM team at IKI acknowledges the subsidy of the Ministry of Science and High Education of Russia.
We warmly thank Michael Smith and another anonymous reviewer for their thorough review of our manuscript.
The processed TIRVIM data and retrievals used in this paper are available in NetCDF format on the Institut Pierre Simon Laplace (IPSL) data server, see \citeA{data}.


%
\newpage
 \bibliography{mars}
%

%
%
%
%
%

\end{document}